\documentclass{jfm}
\usepackage{graphicx}
\usepackage{epstopdf, epsfig}
\usepackage{amsmath}
\usepackage{xcolor}

\newif\ifmarkedup
\markedupfalse
\ifmarkedup
\newcommand{\Revision}[2]{\replaced{#2}{#1}}
\else
\newcommand{\Revision}[2]{{\color{black}#2}}
\fi

\newif\ifmarkedupTwo
\markedupTwofalse
\ifmarkedupTwo

\else

\fi

\shorttitle{Mixing of active scalars due to  random  shock waves in two dimensions}
\shortauthor{Joaquim P Jossy and Prateek Gupta }

\title{\Revision{}{Mixing of active scalars due to random shock waves in two dimensions}}

\author{Joaquim P. Jossy
  \and Prateek Gupta \corresp{\email{prgupta@iitd.ac.in}}}

\affiliation{Department of Applied Mechanics, Indian Institute of Technology Delhi, New Delhi 110016, India.}

\begin{document}

\maketitle

\begin{abstract}

In this work, we investigate the mixing of active scalars in two dimensions by the stirring action of stochastically generated shock waves. We use direct numerical simulations (DNS) of the interaction of shock waves with two non-reacting species to analyse the mixing dynamics for different Atwood numbers ($At$). Unlike passive scalars, the presence of density gradients in active scalars \Revision{alters the molecular diffusion term}{makes the species diffusion nonlinear, introducing} a concentration gradient-driven term and a density gradient-driven \Revision{}{nonlinear} dissipation term in the concentration evolution equation. We show that the direction of concentration gradient causes the interface across which molecular diffusion occurs to expand outward or inward, even without any stirring action. Shock waves enhance the mixing process by increasing the perimeter of the interface and by sustaining concentration gradients. Negative Atwood number mixtures sustain concentration gradients for longer time than positive Atwood number mixtures due to the so-called \Revision{}{nonlinear} dissipation terms. We estimate the time till when the action of stirring is dominant over molecular mixing. We also highlight the role of baroclinicity in increasing the interface perimeter in the stirring dominant regime. We compare the stirring effect of shock waves on mixing of passive scalars with active scalars and show that the vorticity generated by baroclinicity is responsible for the folding and stretching of the interface in the case of active scalars. We conclude by showing that lighter mixtures with denser inhomogeneities ($At<0$) take longer time to homogenise than the denser mixtures with lighter inhomogeneities ($At>0$).
\end{abstract}

% \begin{keywords}
% 
% \end{keywords}
\section{Introduction}
 Mixing is the process of evolution of constituent concentrations from an initial state of segregation to a final state of homogeneity~\citep{villermaux2019mixing}. Usually, the influence of turbulence on mixing of passive scalars is studied to quantify turbulent mixing~\citep{sreenivasan2019turbulent}. The evolution of the concentration ($c$) of a passive scalar is governed by the advection and molecular diffusion terms as shown in \eqref{eq: Passive scalar evolution field}. Molecular diffusion acts across the interface and reduces the mean value of concentration gradients~\citep{eckart1948analysis}. In the absence of advection, mixing by molecular diffusion occurs at a relatively slow rate~\citep{tennekes1972first}. Any action that affects the advective term of \eqref{eq: Passive scalar evolution field} is termed as `stirring'. Stirring enhances the mixing process by increasing the mean value of gradients and increases the area across which molecular diffusion occurs~\citep{eckart1948analysis}. Mixing comprising of both stirring and molecular diffusion can be split into two regimes - the regime where stirring dominates over molecular diffusion and the regime where molecular diffusion becomes the significant contributor to mixing~\citep{meunier2003vortices,sreenivasan2019turbulent}. In this work, we investigate the stirring effect of nonlinear acoustic waves (weak shock waves) and analyse the mechanism by which shock waves enhance the mixing process.
 
\begin{equation}
    \frac{\partial c}{\partial t} +\boldsymbol{u} \bcdot \bnabla c = D \bnabla^{2} c.
    \label{eq: Passive scalar evolution field}
\end{equation}

~\citet{dimotakis2005turbulent} classifies the mixing process into three levels based on the effect of the mixing constituents on the flow dynamics. The mixing of passive scalars is called level-1 mixing, where the mixing constituents have no effect on the flow dynamics \Revision{}{irrespective of the coupling due to density gradients}. Examples involve the mixing of pollutants, dyes, smoke particles, etc, in fluid. The mixing of active scalars is termed as level-2 mixing, where the mixing process is coupled with the flow dynamics through density gradients. Rayleigh-Taylor and Ritchymer-Meshkov instabilities (RMI)~\citep{richtmyer1954taylor,meshkov1969instability} are examples of level-2 mixing. Mixing that proceeds with changes in fluid composition is called as level-3 mixing. Combustion and detonations are some of the best examples of level-3 mixing. There exists an abundant amount of literature on the mixing of passive scalars in both incompressible and compressible flow fields. Experimental studies on the mixing of passive scalars show that scalar gradients decay with time, and the role of vorticity in turbulent fields is to sustain scalar gradients~\citep{buch1996experimental}. \citet{meunier2003vortices} study the mixing of a passive scalar blob using a diffusing Lamb–Oseen type vortex, and show that the changes in concentration can be used to indicate the time in which the effect of stirring is dominant. They identify `mixing time' as an indicator to denote the time during which the action of stirring is dominant over molecular diffusion. They show that the maximum concentration grows linearly till mixing time is reached, after which it decays in time as $t^{-3/2}$. \citet{schumacher2005statistics} derive a relation between the geometrical properties and the mixing statistics of passive scalars using direct numerical simulations (DNS) of a stochastically forced isotropic turbulent field. They relate the area-volume ratio to the passive scalar concentration fluctuations and show that the fluctuations mostly follow a Gaussian distribution. \citet{ni2016compressible} reports that the convolution of scalar fields in forced compressible turbulent flow fields is more pronounced in forcing schemes with a higher ratio of solenoidal to dilational components. In a similar study, \citet{john2019solenoidal} show that the breakdown of the passive scalar field depends on the solenoidal components and that the role of compressibility is negligible in the mixing of passive scalars. \citet{gao2020parametric} challenge the idea of negligible compressibility effects on mixing by showing the alignment of shock with scalar gradients affects mixing in a canonical shock-turbulence interaction configuration. \Revision{reviewer 3}{  However, only a few studies exist that analyse the level-2 and level-3 mixing in compressible flows. In this work, we analyse the level-2 mixing of two non-reacting fluids of different densities and study the effect of density inhomogeneity on mixing and flow dynamics in a randomly forced shock-wave field in a two dimensional setup.
}

RMI refers to the growth of perturbations in the interface between two fluids of different densities under the influence of an impulsive acceleration. RMI is observed in various man-made applications such as \Revision{}{inertial-confinement fusion and supersonic combustion}~\citep{thomas2012drive,yang1993applications}, \Revision{}{as well as naturally occurring phenomena in astrophysics}~\citep{zhou2017rayleigh, burrows2000supernova,arnett2000role}. In cases where shock waves are used for impulsively accelerating the interface, the baroclinic vorticity generated at the interface by the misalignment of pressure gradients and density gradients triggers the growth of perturbations on the interface.  The growth of these perturbations triggers other instabilities like Kelvin-Helmholtz instabilities, leading to enhanced mixing~\citep{brouillette2002richtmyer}. RMI enhanced mixing plays an important role in the design of inertial-confinement fusion reactors~\citep{thomas2012drive}, and the combustion chambers of supersonic air-breathing engines~\citep{yang1993applications}. The change in interface profile caused by RMI is used to develop stellar evolution models to explain the evolution of supernovae explosions~\citep{burrows2000supernova,arnett2000role}. RMI is also the fundamental process for studying shock-flame interactions, where the density difference between the two species is used to model burnt and unburnt fuel mixtures, and the passage of shock waves is used to study the flame propagation and the formation of detonation waves \citep{al2020modeling}. Primarily, the interfacial mixing and the dynamics of the interface are studied in RMI. \citet{yu2021scaling} show that density gradients across the interface govern the mixing rate. They derive an exact mixing rate expression and show the difference in mixing rates during different stages of RMI. \citet{gao2016formula} use Favre averaging of mass fraction to derive the growth rate of the mixing zone width and show that compressibility suppresses the mixing rate for a single pass RMI case. They also report a reduction in the role of the diffusion term in mixing with time. The interface dynamics in RMI are explained in terms of the kinematics of the large scale coherent structures -- spikes (penetration of denser fluid into lighter fluid) and bubbles (penetration of lighter fluid into denser fluid). \citet{zhang1998analytical} uses the Layzer-type model (infinite density ratio across the interface) to show that spikes and bubbles have different growth rates.  \citet{herrmann2008nonlinear} report that bubbles across interfaces of similar density differences move faster and flatten slower compared to interfaces of higher density differences. They conclude that the evolution of the interface is a non-local and multiscale process.  \citet{li2021growth} develop an analytical model to explain the compression and stretching effect of mixing layer for a single passage of shock wave. They also show that the mixing triggered by RMI lasts even after the passage of the shockwave. In the present work, we use stochastically forced shock waves to induce multi-modal perturbations on the interface. We study the mixing induced by such shock waves across interfaces of different density ratios. \Revision{3}{To this end,  we use a simplified two-dimensional model  to analyse the mixing characteristics and interface dynamics resulting from continuous and stochastic forcing of a setup resembling that of RMI using shock-resolved direct numerical simulations (DNS) of the fully compressible Navier-Stokes equations.} 

In this work, we analyse the mixing of active scalars driven by random shock waves. We consider \Revision{}{binary mixtures of }two non-reacting species with different density ratios. Throughout, we consider a circular shaped inhomogeneity (blob) with a smaller area than the rest of the domain. In \S\ref{sec:Governing Equation and Numerical Simulations}, we discuss the governing equations we use to study the mixing enhancement effect of shock waves. We also discuss the details of the numerical setup and the parameters used to carry out the shock-resolved DNS. In \S\ref{sec:Results and Discussion}, we first discuss the effect of density gradients on the diffusion of active scalar \Revision{}{without shock waves}. We use a \Revision{}{simplified }one-dimensional \Revision{set-up}{nonlinear diffusion equation} to show that the direction of the density gradients affects the evolution of mass fraction \Revision{}{due to nonlinear diffusion}. We next show the stirring effect of shock waves and its role in enhancing the mixing. We also show the role of baroclinicity in increasing the perimeter across which diffusion occurs before concluding with a summary of our findings in  \S\ref{sec: Conclusions}.

\section{Governing Equations and Numerical Simulations}
\label{sec:Governing Equation and Numerical Simulations}
In this section, we discuss the \Revision{theoretical background}{governing equations} and the numerical method used to study the mixing of two non-reacting species by stochastically forced shock waves. We use the previously developed forcing scheme for generating a random shock-wave field~\citep{jossy2023baroclinic}  and summarize it below along with the parameter space.

We use the dimensionless form of the Navier-Stokes and species equation to analyse the mixing of active scalars by shock waves. A non-reacting binary mixture is used to create different density inhomogeneities by varying the molecular weight ($W_{i}$)  of the species. The dimensionless equations of the density $\rho$, velocity $\boldsymbol{u}$, pressure $p$, and concentration of the $i^{\mathrm{th}}$ species $Y_i$ (mass fraction) are given by,

\begin{equation}
    \frac{\partial \rho}{ \partial t} +  \bnabla\bcdot(\rho \boldsymbol{u}) =0 ,
    \label{eq: continuity}
\end{equation}

\begin{equation}
    \frac{\partial \boldsymbol{u} }{\partial t} + \boldsymbol{u} \bcdot \bnabla \boldsymbol{u} + \frac{\bnabla p}{\rho} = \frac{1}{ \rho \mathrm{Re}_{\mathrm{ac}} }  \left( \bnabla\bcdot( 2 \mu \boldsymbol{S} )\right) +
  \frac{1}{ \rho \mathrm{Re}_{\mathrm{ac}} }  \left( \bnabla\bcdot \left( \kappa - \frac{2 \mu }{3} \right) \boldsymbol{D} \right) + \boldsymbol{F} ,
  \label{eq: momentum_equation} 
\end{equation}

\begin{equation}
    \frac{\partial p}{\partial t} + \boldsymbol{u} \bcdot \bnabla p + \gamma p\bnabla \bcdot \boldsymbol{u} =  \frac{1}{\mathrm{Re}_{\mathrm{ac}} \hspace{0.8mm} \mathrm{Pr}} \bnabla \bcdot (\alpha \bnabla T) + \frac{\gamma -1 }{\mathrm{Re}_{\mathrm{ac}}}\left(2 \mu \boldsymbol{S}:\boldsymbol{S} +\left( \kappa-\frac{2 \mu}{3}\right)\boldsymbol{D}:\boldsymbol{D}   \right) ,
 \label{eq: pressure_equation}
\end{equation}

\begin{equation}
    \frac{\partial Y_{i}}{\partial t} + \boldsymbol{u} \bcdot \bnabla Y_{i} = \frac{1}{\rho \hspace{0.8mm} \mathrm{Re}_{\mathrm{ac}}  \mathrm{ Sc}\hspace{0.8mm} } \bnabla \bcdot \left( \rho \frac{W_{i}}{W} \bnabla X_{i}\right),
     \label{eq: species equation}
\end{equation}
where $W$ is the mean molecular mass of the mixture, and $X_i$ is the mole fraction of the ${i}^{\mathrm{th}}$ species related to the mass fraction $Y_i$ as, 
\begin{equation}
     X_{i} = \frac{W}{W_{i}} Y_{i} .
        \label{eq: molar mass equation}
\end{equation}
We combine the above equations using the dimensionless equation of state for an ideal gas mixture, 
\begin{equation}
\gamma p = \frac{\rho}{W} T,\label{eq: ideal_gas_nd}        
\end{equation}
\Revision{}{where}
\begin{equation}
    \frac{1}{W} = \frac{Y_{1}}{W_{1}} + \frac{Y_{2}}{W_{2}}.
        \label{eq: mean molecular weight equation}
\end{equation}
\Revision{}{Equation~\eqref{eq: pressure_equation} is derived using the internal energy equation for an ideal gas~\citep{kundu2015fluid}.} 
In \eqref{eq: momentum_equation} and \eqref{eq: pressure_equation}, $\boldsymbol{S}$ and $\boldsymbol{D}$ are the strain rate and dilatation tensors, respectively, given by, 
\begin{equation}
 \boldsymbol{S} = \frac{1}{2}\left(\bnabla\boldsymbol{u} + \bnabla\boldsymbol{u}^T\right), ~\boldsymbol{D} = \left(\bnabla\bcdot\boldsymbol{u} \right) \boldsymbol{I} .
\end{equation}
\Revision{}{The fluid properties such as dynamic viscosity ($\mu$), bulk viscosity ($\kappa$), and thermal conductivity ($\alpha$) are all dimensionless. The flow properties are governed by the dimensionless numbers, the acoustic Reynolds number ($\mathrm{Re}_{\mathrm{ac}}$), the Prandlt number ($\mathrm{Pr}$), and the Schmidt number ($\mathrm{Sc}$) defined as,
\Revision{point 4}{\begin{equation}
		\mathrm{Re}_{\mathrm{ac}} = \frac{\rho_{m} c_{m} L_{m}} {\mu_{m}}, \hspace{2 mm} \mathrm{Pr} = \frac{ \gamma \mu_{m} \tilde{R} }{(\gamma - 1) W_{m} \alpha_{m}},\hspace{2 mm} \mathrm{Sc} = \frac{\mu_{m}}{\rho_{m} \lambda_{m}}.
		\label{eq: Non-dimesnaional parameters}
\end{equation}}
The flow field parameters are scaled using the characteristic dimensional quantities which are denoted by the subscript $()_{m}$. The velocity is scaled using the characteristic speed of sound $(c_{m})$, and $\tilde{R}$ $\mu_{m}, \alpha_{m} $, and $\lambda_{m}$ denote the universal gas constant, dimensional viscosity, thermal conductivity, and  diffusion coefficient respectively. Since we use dimensionless equations in our numerical simulations, the  specific values of the characteristic scales hold no significance. We also assume that the heat capacities of both the species are the same.}

We refer the reader to our earlier work \citep{jossy2023baroclinic} for a detailed derivation of the dimensionless governing equations~\eqref{eq: continuity}-\eqref{eq: pressure_equation} and the appropriate characteristic scales of the quantities. The forcing term $\boldsymbol{F}$ in \eqref{eq: momentum_equation} is used to generate shock waves.  Since shock waves generate intense pressure gradients, we use the Hirschfelder and Curtiss approximation~\citep{hirschfelder1964molecular,poinsot2005theoretical} for the species diffusion in \eqref{eq: species equation}. We highlight that when pressure gradients are small, the diffusion term in \eqref{eq: species equation} becomes the right-hand side (RHS) of \eqref{eq: Passive scalar evolution field}. \Revision{}{As we discuss in later sections, the nonlinear diffusion due to Hirschfelder and Curtiss approximation causes the active scalar interface diffusion to travel, resulting in an expansion or contraction of the inhomogeneities, depending on the value of the molecular mass ratio.}

% We scale pressure, velocities, temperature, and the coordinates in all the above equations with their corresponding characteristic quantities following \citet{jossy2023baroclinic}. The dimensionless quantities are also calculated based on these characteristic quantities.
\begin{subsection}{Numerical Simulations}
\Revision{}{We perform shock-resolved direct numerical simulations (DNS) of the fully compressible two-dimensional Navier-Stokes equations \eqref{eq: continuity} - \eqref{eq: pressure_equation} and the species equation \eqref{eq: species equation} using an MPI parallelised pseudospectral solver}. We use the Runge-Kutta fourth-order scheme for time stepping and ensure numerical stability using the 2/3 deliasing method \Revision{}{in the nonlinear terms only to eliminate the aliasing errors. As we show later, nonlinearities are negligible for length scales much larger than the length scale below which the aliasing errors are removed}. For all our simulations, we consider 3072 Fourier modes in each direction such that \Revision{}{the nonlinear terms are resolved till $k_{\mathrm{max}}=1024$.} We show in \S \ref{Parameters} that for the energy injection rates considered~\Revision{}{in this work ($\varepsilon$ in \eqref{eq: forcing_injection})}, the resolution is sufficient to resolve the smallest length scales (both the shock thickness and Batchelor scales).  Each simulation is run using 256 cores consuming approximately $10^{5}$ core hours. 
We consider constant viscosity ($\mu =1 $) and thermal conductivity ($\alpha =1 $) for all our simulations and ignore bulk viscosity ($\kappa = 0$) for simplicity. For all simulations, we consider $\mathrm{Pr} = \mathrm{Sc} = 0.7$, such that the molecular diffusion and heat diffusivity remain the same.
We initialise our simulations using quiescent ($\boldsymbol{u} =0 $), isobaric (uniform pressure), and isothermal (uniform temperature) conditions~\Revision{}{with a circular blob of radius $\pi/2$.} We \Revision{specify the  species having a circular distribution  using a tanh  function as follows}{use the $\tanh$ function to specify the initial concentration field as follows}
\begin{equation}
     Y_{c} =\frac{1}{2} \left[1 - \mathrm{tanh}\left( \frac{1}{\delta} \left (\sqrt{ (x - \pi)^{2} + (y - \pi)^2} - \frac{\pi}{2} \right )\right) \right] .
    \label{eq: circular species ditribution}
\end{equation} The species constituting the blob has density $\rho_{c}$ and the surrounding species has density $\rho_{s}$. We use subscript $()_c$ to indicate the species inside the circular blob and the subscript $()_s$ to denote the surrounding species. We use the \Revision{}{parameter value $\delta = 1/15$} in \eqref{eq: circular species ditribution} to obtain a steep but smooth concentration distribution to achieve a \Revision{sharp but diffused interface}{diffused thin interface}. The parameter $\delta$ represents a characteristic interface thickness and will be used later to calculate the apparent perimeter of the interface~\Revision{}{during mixing}.

\Revision{3}{The forcing $\boldsymbol{F}$ in \eqref{eq: momentum_equation} is similar to the one used by \cite{jossy2023baroclinic}. For completeness, we provide a brief overview here. The forcing $\boldsymbol{F}$ is defined in the Fourier space and computed using four statistically independent Uhlenbeck-Ornstein (UO) \citep{eswaran1988examination}, processes ($a_{i,\boldsymbol{k}} (t)$ for $i = 1,2,3$ and $4$),  satisfying the following conditions
	\begin{align}
		&\left\langle a_{i, \boldsymbol{k}}(t)\right\rangle = 0 ,\\
		&\left\langle {a}_{i, \boldsymbol{k}}(t) {a}_{j, \boldsymbol{k}}^{*}(t+s)\right\rangle = 2 \sigma^{2} \boldsymbol{\delta_{ij}}\exp(-s/T_L) ,
	\end{align}
	where $\left\langle\cdot\right\rangle$ denotes the ensemble average, $()^*$ denotes the complex conjugate, $\sigma^2$ is the variance, and $T_L$ is the forcing time-scale.} To prevent any direct effect of energy injection at small length scales, we restrict our forcing to larger scales within the band of wavenumbers $0  < |\boldsymbol{k}| < k_{F} $, where $k_{F} = 5 $ in all our simulations. The  total rate of energy injection $\varepsilon$ is a function of the variance $\sigma^{2}$, time scale $T_{L}$ of the UO process, and the number of wave vectors within the forcing band $N_{F}$ and is given by 
\begin{equation}
    \varepsilon = 4 N_{F} T_{L} \sigma^{2}.
    \label{eq: forcing_injection}
\end{equation}
\Revision{3}{The forcing vector in the Fourier space can be obtained by combining the four independent processes as, 
	\begin{equation}
		\widehat{\boldsymbol{a}}_{\boldsymbol{k}} = \left(a_{1,\boldsymbol{k}} + i a_{2,\boldsymbol{k}}, a_{3,\boldsymbol{k}} + ia_{4,\boldsymbol{k}}\right)^T.
		\label{eq: randomForcingVector}
	\end{equation}
To force only the acoustical component of velocity, we remove the vortical component from $\widehat{\boldsymbol{a}}_{\boldsymbol{k}}$ yielding
\begin{align}
	\widehat{\boldsymbol{F}}_{\boldsymbol{k}}(t) = \frac{(\widehat{\boldsymbol{a}}_{\boldsymbol{k}}(t) \cdot\boldsymbol{k})}{|\boldsymbol{k}|^2}\boldsymbol{k}.
 \label{eq: forcing_final}
\end{align} 
}
\Revision{3}{It has been shown that decaying acoustic wave turbulence has similar behaviour to Burgers turbulence \citep{gupta2018spectral}. We speculate that the behaviour of stochastically forced shock waves may be similar to that of two-dimensional forced Burgers turbulence \citep{cho2014statistical}. A detailed comparison between the two is beyond the scope of the current study. In our earlier work~\citep{jossy2023baroclinic}, we have used the stochastic forcing method described above to generate a field of isotropic shock waves. The wave energy spectra of the random shock waves generated exhibit identical dependence on the dissipation and the integral length scale as decaying one-dimensional turbulence \citep{gupta2018spectral} as well as shallow water wave turbulence~\citep{Augier_Mohanan_Lindborg_2019}. \citet{Augier_Mohanan_Lindborg_2019} derive the shallow water wave-turbulence scaling laws assuming isotropy, independent of the dispersive nature of the shallow water waves. Hence, the forcing method mentioned in~\eqref{eq: forcing_final} generates a field of isotropic random shock waves whose phasing changes at every time instant in a homogeneous mixture. Reflections of the shock waves at the inhomogeneity interface may introduce some spatial anistropy, a detailed analysis of which is deferred to future studies and is beyond the scope of the current work.} As mentioned earlier, in this work, we study the interaction of \Revision{}{forced }shock waves, which induce multimodal perturbations on the \Revision{}{inhomogeneity} interface separating the two species and thus influencing the active scalar mixing \Revision{process}{dynamics}.   

\end{subsection}
\begin{subsection}{Simulation Parameters}
\label{Parameters}
\begin{table}
  \begin{center}
\def~{\hphantom{0}}
  \begin{tabular}{lcccc}
      Sl.no  & Active/Passive   &   $At$ & $\chi$  & Indicator\\
       1   & Passive & ~~--~ & --&~~ $\mathrm{PS}^*$\\
       2   & Active & ~~0.75 & 7& ~~$At = 0.75^{*}$\\
       3  & Active & ~~-0.75 & 0.1428&~~ $At = -0.75^{*}$\\
  \end{tabular}
  \caption{Parameter space for simulations considering only the effect of molecular diffusion without any shock waves along with their indicators.}
  \label{tab: No shock case matrix}
  \end{center}
\end{table}
\begin{table}
  \begin{center}
\def~{\hphantom{0}}
  \begin{tabular}{lccccccc}
      Sl.no  & Active/Passive   &   $At$ & $\chi$ & \Revision{}{$\langle M \rangle$}  &\Revision{}{$\langle  \eta \rangle_{t}$} & \Revision{}{$ \langle \eta_{B} \rangle_{t}/\Delta$} &Indicator\\
       1   & Passive & ~~--~ & --&\Revision{}{1.09}&\Revision{}{0.030}&\Revision{}{12.2} &~~ PS\\
       2   &\Revision{}{Active}  & \Revision{}{0.25}  &\Revision{}{1.67} & \Revision{}{1.08}&\Revision{}{0.038}&
       \Revision{}{15.8} &~~~~\Revision{}{$At = 0.25^{\dag}$}\\
       3  & Active & 0.5 & 3.0& \Revision{}{1.07}&\Revision{}{0.023}&\Revision{}{9.3} &~~ $At = 0.5$\\
       4   &\Revision{}{Active}  & \Revision{}{0.75} & \Revision{}{7.0} & \Revision{}{1.07}& \Revision{}{0.019}&\Revision{}{7.7} &~~~ \Revision{}{$At = 0.75^{\dag}$}\\
       5   & \Revision{}{Active} &\Revision{}{ -0.25} & \Revision{}{0.6}& \Revision{}{1.08}& \Revision{}{0.029}&
       \Revision{}{11.8} &~~~~~\Revision{}{$At = -0.25^{\dag}$}\\
       6  & Active & -0.5 & 0.33& \Revision{}{1.07}&\Revision{}{0.028}&\Revision{}{11.3} &~~~ $At = -0.5$\\
       7  & \Revision{}{Active} & \Revision{}{-0.75} & \Revision{}{0.1428}& \Revision{}{1.08}&\Revision{}{0.022}&\Revision{}{9.37} &~~~~ \Revision{}{$At = -0.75^{\dag}$}\\
  \end{tabular}
  \caption{Parameter space for simulations with stochastically generated shock waves and their corresponding indicators. \Revision{}{In the above table, $\langle M \rangle$ represents the area averaged Mach number of the shock waves in the domain, $\langle \eta \rangle_{t}$ represents the time-averaged shock thickness scale, and $\langle \eta_{B} \rangle$ represents the time averaged Batchelor scale.} }
  \label{tab: shock wave parameter space}
  \end{center}
\end{table}
    To study the effect of \Revision{}{forced} shock waves on the mixing of species with different densities, we use Atwood number ($At$) defined as 
    \begin{equation}
        At = \frac{\rho_{s} - \rho_{c}}{\rho_{s} + \rho_{c}} \hspace{1 mm} = \frac{\chi - 1}{\chi + 1} .
    \end{equation}
We study the effect of shock waves in both a denser mixture ($\chi > 1$) and a lighter mixture ($\chi < 1$) where $\chi$ is the density ratio of surrounding species to the blob species. To understand the effect of density gradients on species diffusion, we run one case each of positive and negative Atwood numbers and compare them with a passive scalar case. Table \ref{tab: No shock case matrix} shows the parameters of the simulation cases without any shock waves. \Revision{}{ We study the effect of shock waves on mixing for three values of positive and negative Atwood numbers each and compare their mixing dynamics with a passive scalar case. Since each realisation of the stochastic forcing $\boldsymbol{F}$ is different, the mixing process itself is a stochastic non-equilibrium process. Hence, we simulate additional realisations of the end ranges of the Atwood number ie, 4 realisations of $At = 0.75$, 4 realisations of $At = -0.75$, 3 realisations of $At = 0.25$ and 3 realisations of $At =- 0.25$ to show that the mixing characteristics are consistent across Atwood numbers for different realisations of the same forcing parameters. The average of the different realisations will be represented by the superscript $()^{\dag}$. Table \ref{tab: shock wave parameter space} shows the simulation cases used to study the stirring effect of shock waves. For all simulations, we inject acoustic energy at the rate of $\varepsilon/ 4 N_{F} =1.25 \times 10^{-6} $ to generate weak shock waves ($ \langle M \rangle < 1.1$). We run all the realisations for all the cases for sufficiently long times till the effects of molecular diffusion take over the effects of stirring. We run one realisation for all the different Atwood numbers shown in table \ref{tab: shock wave parameter space} further close to homogenisation.} \Revision{2}{The smallest length scales generated in the simulations correspond to the shock thickness scale $\eta$~\citep{larsson2009direct,donzis2012shock, gupta2018spectral, jossy2023baroclinic} and the Batchelor scale $\eta_B$, which are related to the acoustic Reynolds number $\mathrm{Re}_{ac}$ as,
\begin{equation}
    \eta \sim \frac{1}{\mathrm{Re}_{\mathrm{ac}}} ( \varepsilon \ell)^{-1/3} , \hspace{1 mm} \eta_{B} = \frac{\eta}{\sqrt{\mathrm{Sc}}},
\end{equation}
where $\ell = 2/k_F$ is the integral length scale, which represents the mean distance between two shocks. To ensure shock-resolved simulations, we consider $\mathrm{Re}_{\mathrm{ac}} = 1000$, such that $ k_{\mathrm{max}} \eta > 1.5$ and follow an even more strict resolution scale for mixing, $   \eta_{B}/\Delta > 0.5 $~\citep{schumacher2005very}
where $\Delta$ is given by $ 2\sqrt{2} \pi / 3k_{\mathrm{max}}  $.}
\begin{figure}
\centering
{\includegraphics[width=1.0\textwidth]{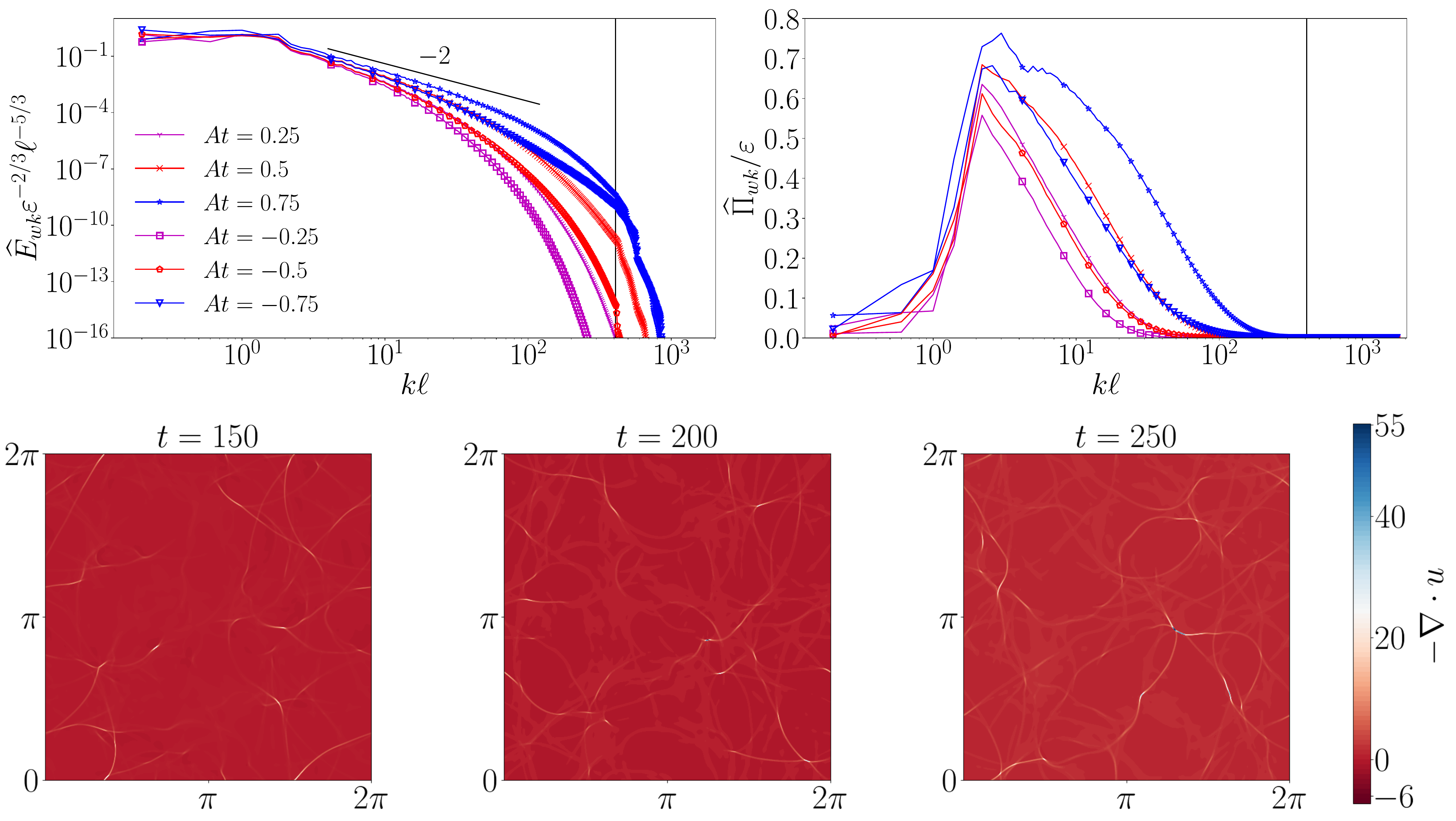}\put(-385,220){(a)}
\put(-200,220){(b)} \put(-385,115){(c)}}
  \caption{\Revision{reviwer 2}{Dimensionless scaled density weighted wave energy $\hat{E}_{wk}$ spectra (a) and spectral flux of wave energy $\widehat{\Pi}_{wk}$ (b)  of active scalars cases in table \ref{tab: shock wave parameter space} averaged over the time interval $t=50$ to $t=300$. The vertical lines in both the figures show the dealiasing limit $k_{\mathrm{max}}\ell$. The spectral flux is negligible beyond the $k_{\mathrm{max}}\ell$ highlighting that all the nonlinear scales are resolved. The legends are identical for both the figures. Contours of divergence at different time instances for $At=0.75$ (c).}}
\label{fig: Energy Spectra}
\end{figure}

\Revision{point 4}{

To show that all our simulations are resolved, we calculate the binned density-weighted wave spectral energy $\widehat{E}_{wk}$ and the spectral flux of the wave energy $\widehat{\Pi}_{wk}$ whose definitions are given in \eqref{eq: Wave Energy spectra} and \eqref{eq: transfer_function_f} respectively. We direct the reader to previous studies ( \citet{miura1995acoustic} and \citet{jossy2023baroclinic}) for the detailed derivation of the density-weighted spectral quantities. We use a density weighted velocity $\boldsymbol{w} = \sqrt{\rho} \boldsymbol{u}$ and its wave component obtained upon the Helmholtz decomposition~\citep{miura1995acoustic,jossy2023baroclinic}. Denoting the Fourier transform of a quantity $\phi$ by $\hat{\phi}$ and the complex conjugate by $\hat{\phi}^{*}$, the density-weighted wave spectral energy $\widehat{E}_{wk}$ and its transfer function $\widehat{\mathcal{T}}_{wk}$ are obtained as,
	\begin{equation}
 \widehat{E}_{wk} = \frac{1}{2} \left( \widehat{\boldsymbol{w}}^{*}_{wk} \cdot \widehat{\boldsymbol{w}}_{wk} + \widehat{\left(p - \frac{1}{\gamma}\right)}_{k}^{*} \widehat{\left(p - \frac{1}{\gamma}\right)}_{k} \right)\label{eq: Wave Energy spectra}
 \end{equation}
 \begin{align}
		&\widehat{\mathcal{T}}_{wk} =\mathcal{R}\left[\widehat{ \boldsymbol{w}}^*_{wk}\cdot\left(\widehat{\frac{\nabla p}{\sqrt{\rho}}}\right)_{k}  + \widehat{\left(p - \frac{1}{\gamma}\right)}^*_{k}\widehat{\nabla\cdot\boldsymbol{u}}_{wk}\right] \nonumber \\ 
		&+\mathcal{R}\left[\hat{\boldsymbol{w}}^*_{wk}\cdot\left(\widehat{\boldsymbol{w}\cdot\nabla\boldsymbol{u}}\right)_{wk} - \widehat{\boldsymbol{w}}^*_{wk}\cdot\left(\widehat{\frac{\left(p - \frac{1}{\gamma}\right)\nabla p}{\sqrt{\rho}}}\right)_{k}\right] \nonumber \\   &+ \mathcal{R}\left[\gamma\widehat{\left(p - \frac{1}{\gamma}\right)}^*_{k}\left(\widehat{\left(p - \frac{1}{\gamma}\right)\nabla\cdot\boldsymbol{u}}\right)_{wk} + \widehat{\left(p - \frac{1}{\gamma}\right)}^*_{k}\left(\widehat{\boldsymbol{u}\cdot\nabla p}\right)_{wk}\right],
		\label{eq: transfer_function_f}
\end{align}
where the subscript $()_{wk}$ denotes the Fourier transform of the wave-field, and $\mathcal{R}()$ denotes the real part of the quantity. Equations~\eqref{eq: Wave Energy spectra} and \eqref{eq: transfer_function_f} are derived using the second order approximation of \eqref{eq: continuity}-\eqref{eq: pressure_equation} (see \cite{jossy2023baroclinic} for a detailed derivation). The term $p - \frac{1}{\gamma}$ is the change in pressure from the isobaric base state at the dimensionless pressure of $1/\gamma$.}
\Revision{}{Figure~\ref{fig: Energy Spectra}(a) shows the scaled density-weighted wave energy spectra $\widehat{E}_{wk}\varepsilon^{-2/3}\ell^{-5/3}$ vs the scaled wavenumber $k\ell$. Since the forcing acts only within $0 < k\ell < 2$, wave energy in the further wavenumbers indicates forward spectral wave energy cascade due to the nonlinear terms in the governing equations. Since the generated shock waves are weak, $\widehat{E}_{wk}$ decays as $k^{-2}$ before decaying rapidly due to thermoviscous dissipation. Figure~\ref{fig: Energy Spectra}(b) shows the scaled flux of the spectral wave energy $\widehat{\Pi}_{wk}/\varepsilon$ calculated as the cumulative sum of the transfer function $\widehat{\mathcal{T}}_{wk}$ in \eqref{eq: transfer_function_f}. The spectral flux approaches net dissipation (equal to net energy injection) and then decays after a decade as the wavenumber increases. Unlike hydrodynamic turbulence, thermoviscous dissipation is significant at all scales in a field of random shock waves since the wave energy $\sim k^{-2}$ and the dissipation scales as $k^2\widehat{E}_{wk}$ approximately. Similar values of spectral flux and the net energy injection highlight that the waves generated are nonlinear.
Decay of the spectral flux in figure~\ref{fig: Energy Spectra} before $k_{\mathrm{max}}\ell$ for all the cases shows that nonlinearities are not negligible at length scales larger than $k_{\mathrm{max}}$. Since the dealiasing is used only in the nonlinear terms, any linear dynamics at length scales smaller than $1/k_{\mathrm{max}}$ till $1/3072$ are well resolved.

The variation of density-weighted wave spectra across different Atwood number cases occurs due to the variation in density across the inhomogeneity interface. Stretching and folding of the inhomogeneity generate smaller scales in the interface. However, the scale corresponding to the interface thickness increases due to molecular diffusion. As discussed later in \S~\ref{sec:Effect of density gradients on molecular diffusion}, the molecular diffusion term responsible for the diffusion of the interface (later termed as the Laplacian diffusion term) and the molecular diffusion term responsible for the movement of the interface due to the density gradients (later termed as the \Revision{}{nonlinear} dissipation term) change with the Atwood number (c.f.\eqref{eq: D1 coeff}). For $At>0$, the Laplacian diffusion term dominates the \Revision{}{nonlinear} dissipation term. As $At>0$ increases, the Laplacian diffusion coefficient decreases, indicating relatively smaller scales maintained for higher values of $At>0$ in figure~\ref{fig: Energy Spectra}(a). As $At<0$ decreases, the \Revision{}{nonlinear} dissipation coefficient increases resulting in more rapid expansion of the interface. This results in more collisions of the shock waves with the interface generating smaller scales in the interface for more negative $At$ cases. It is important to note that in figure~\ref{fig: Energy Spectra}, the density gradients due to mixture being inhomogeneous have the smallest contribution for $At=\pm 0.25$ cases since the density values of the two species are the closest among the cases considered. }
\Revision{}{	 \begin{figure}
  \centering
\includegraphics[width=1.0\textwidth]
{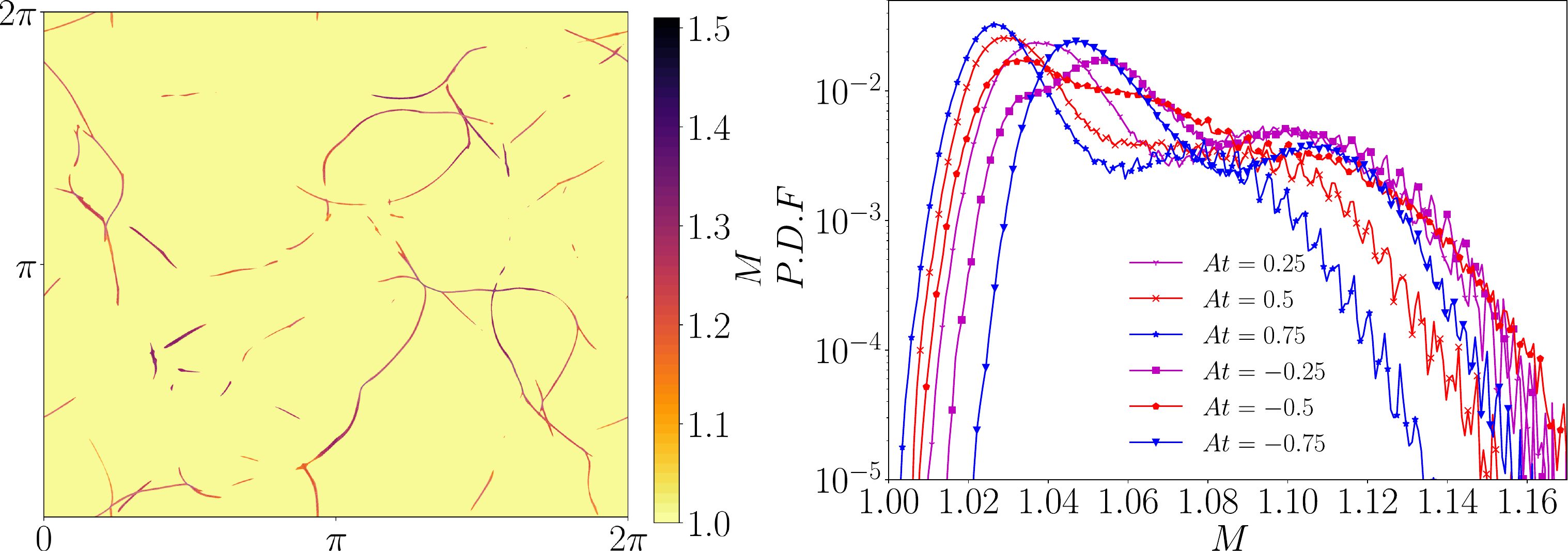}
\put(-385,145){(a)}
\put(-182,145){(b)}
	\caption{ \Revision{}{Mach number contours of $At = 0.75$ at $t= 250$ calculated using \eqref{eq: Mach eqaution} (a).Time averaged normalised histogram (PDF) of the Mach number of active scalars cases in table \ref{tab: shock wave parameter space} (b).}}
	\label{fig: Mach contour and Mach PDF}
\end{figure}
}

\Revision{reviewer 2}{In viscous compressible simulations, shocks can be identified by the negative divergence lines \citep{samtaney2001direct,Augier_Mohanan_Lindborg_2019}. Figure \ref{fig: Energy Spectra}(c) shows the contours of negative divergence for $At = 0.75$ at different time instances. The high negative values of divergence are indicative of the presence of shocks in the field. To quantify the strength of the random shock waves generated, we calculate the shock Mach number using the dimensionless entropy. The entropy for a mixture of species is defined as~\citep{liepmann2001elements} 
	\begin{equation}
		s = \frac{1}{W} \left( \frac{\gamma}{\gamma -1} \log \frac{T}{T_o}\right) - \sum_{i=1}^{2} \frac{1}{W_{i}} \log\frac{p_{i}}{p_{io}}.
		\label{eq: entropy eqaution}
	\end{equation}
where $p_{i}$ is the partial pressure of species $i$ and the subscript $()_{o}$ indicates the initial state of the species. At a point in the domain, entropy changes due to mixing as well as the entropy generated due to shocks. To isolate the entropy generated due to the shock waves only, we calculate the entropy jump only in the regions where the divergence is negative ($\nabla \cdot \boldsymbol{u} < 0$). We calculate the approximate entropy jump using the entropy field $s$ as,
\begin{equation}
    \Delta s \approx \left|\frac{\partial s}{\partial x}\right|dx + \left|\frac{\partial s}{\partial y}\right|dy.
\end{equation}
Using the entropy jump field $\Delta s$, we calculate the Mach number field assuming weak shocks using 
	 \begin{equation}
	 	M = \sqrt{1 + \left(\frac{3\Delta s(\gamma + 1)^2}{2\gamma}\right)^{1/3}}.
	 	\label{eq: Mach eqaution}
	 \end{equation} 
 	
	 Figure \ref{fig: Mach contour and Mach PDF}(a) shows the contours of Mach number for the case $At =0.75$ at one time instant. In figure \ref{fig: Mach contour and Mach PDF}(b), we show the time-averaged normalised histogram or probability distribution function (PDF) of the Mach number field $M$ for all the Atwood number cases with shocks. Since the shock waves are very thin, most of the domain exhibits Mach number very close to 1.  A single value of the representative mean Mach number for all the cases with shocks is shown in table \ref{tab: shock wave parameter space}.
}

\end{subsection}

\vspace{4 mm}
\section{Results and Discussion}
\label{sec:Results and Discussion}
\Revision{1}{In this section, we present the mixing effects of the forced shock waves on active scalars in a two-dimensional setup. Density gradients affect both the advection and diffusion of active scalars.}
 In the presence of shock waves, the baroclinic interaction of density gradients with shock waves results in a vorticity field, which tends to deform the blob and increase the total apparent interface perimeter through advection. Additionally, the density-weighted \Revision{}{nonlinear} diffusion of active scalars is significantly different from the diffusion of passive scalars. We first discuss the mixing of active scalars in the absence of shock waves to isolate the effect of the density gradients on the diffusion process. We then discuss the enhancement of mixing by shock waves and finally discuss the role of baroclinicity in increasing the apparent perimeter of the interface through advection.

\begin{subsection}{Effect of density gradients on \Revision{}{nonlinear} molecular diffusion}
\label{sec:Effect of density gradients on molecular diffusion}
\begin{figure}
 \includegraphics[width=\textwidth]{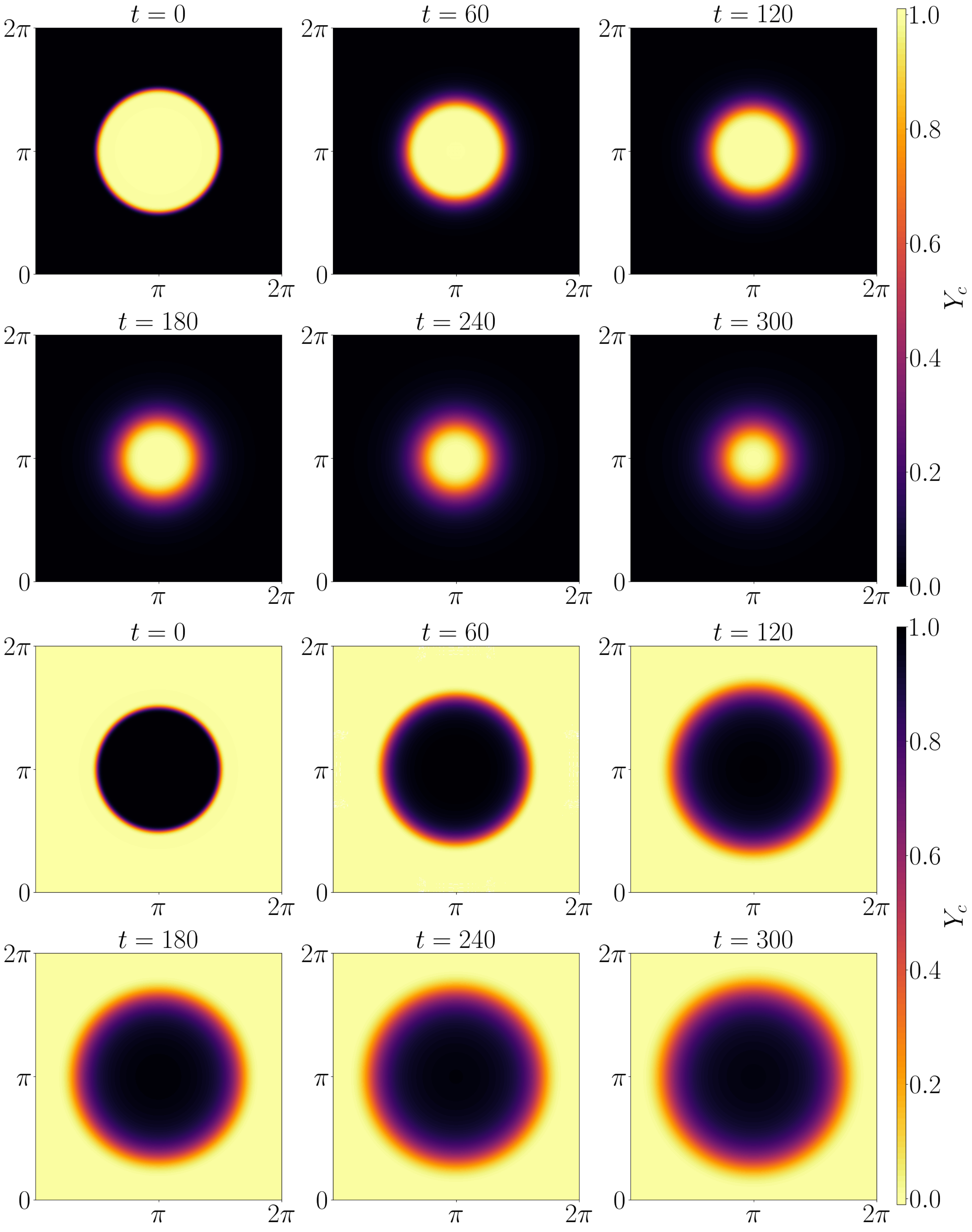}
 \put(-400,480){$(a)$}
 \put(-400,235){$(b)$}
\caption{Contours of concentration of blob for $At = 0.75^{*}$ (a) and $At = -0.75^{*}$ (b) at same times. Darker colour indicates heavier species. The changes in concentration indicate the mixing driven solely by molecular diffusion. The lighter blob shrinks while the heavier blob expands. In the above contours, the lighter colour indicates the species which is less denser while the darker colour indicates the species which is heavier.}
\label{fig:No shock conc contours}
\end{figure}

The role of molecular diffusion in passive scalars is to reduce the mean value of concentration gradients. In this section, we analyse the effect of density gradients on the \Revision{}{nonlinear} molecular diffusion process of active scalars. To isolate the effect of density gradients on mixing by molecular diffusion, we consider cases, $At = 0.75^{*}$ and $At = -0.75^{*}$  where mixing is solely due to molecular diffusion. These cases are compared with the diffusion of a passive scalar case $\mathrm{PS}^*$ in table \ref{tab: No shock case matrix}, where no density gradients are present. In these three cases, $\boldsymbol{F}$ in \eqref{eq: momentum_equation} is set to zero. Without shock waves, the advective term of the concentration evolution equation becomes negligible, and \eqref{eq: species equation} can be approximated as 
\begin{equation}
    \frac{\partial Y_{i}}{\partial t} 
    = \frac{1}{\rho \hspace{0.8mm} \mathrm{Re}_{\mathrm{ac}}  \mathrm{ Sc}\hspace{0.8mm} } \bnabla  \bcdot \left( \rho \frac{W_{i}}{W} \bnabla X_{i}\right).
     \label{eq:pure diffusion  species equation Results}
\end{equation}
  The RHS of \eqref{eq:pure diffusion  species equation Results}  can be written as follows
  \begin{equation}
  \frac{1}{\rho \hspace{0.4mm} \mathrm{Re}_{\mathrm{ac}} \hspace{0.4mm} \mathrm{ Sc} } \bnabla \left( \rho \frac{W_{i}}{W} \bnabla X_{i}\right)   = \frac{1}{\rho \hspace{0.4mm} \mathrm{Re}_{\mathrm{ac}} \hspace{0.4mm} \mathrm{ Sc}} \bnabla \bcdot \left( \rho \frac{\bnabla W}{W} Y_{i} +\rho \bnabla Y_{i}\right).
       \label{eq: Species diffusion broken down Results}
  \end{equation}
 The RHS  of \eqref{eq: Species diffusion broken down Results} represents the diffusion of an active scalar. Unlike the \Revision{}{linear} diffusion of a passive scalar (c.f. \eqref{eq: Passive scalar evolution field}), which depends only on the concentration gradients, the diffusion of an active scalar also depends on the mean molecular mass gradients \Revision{}{and is nonlinear}. Using the definition of mean molecular mass and the mass fraction conservation constraint ($Y_c + Y_s = 1$), the equation for the evolution of the circular species ($Y_{c}$) can be written in terms of gradients of species concentration as, 
 \begin{equation}
 \frac{\partial Y_{c}}{\partial t} =\frac{1}{ \mathrm{Re}_{\mathrm{ac}} \hspace{0.1mm} \mathrm{ Sc}}\left(
    \underbrace{ D_1 \bnabla^2 Y_{c}}_{I}
    +\underbrace{ D_2 \bnabla Y_{c} \bcdot\bnabla Y_{c}}_{II}
   + \underbrace{\frac{D_1}{\rho}\left(  \bnabla \rho \bcdot \bnabla Y_{c}\right) }_{III}\right).
   \label{eq: Pure Diffusion equation expanded}
 \end{equation}
where $D_1$ and $D_2$ are defined as 
 \begin{equation}
       D_{1} = \frac{1}{1+ (\chi - 1)Y_{c}}, \hspace{2mm} D_{2} = \frac{ 1 - \chi}{\left(1+ (\chi - 1)Y_{c}\right)^{2}} \hspace{0.8mm}.
       \label{eq: D1 coeff}
   \end{equation}
   
\begin{figure}\centering
\includegraphics[width=1.0\textwidth]{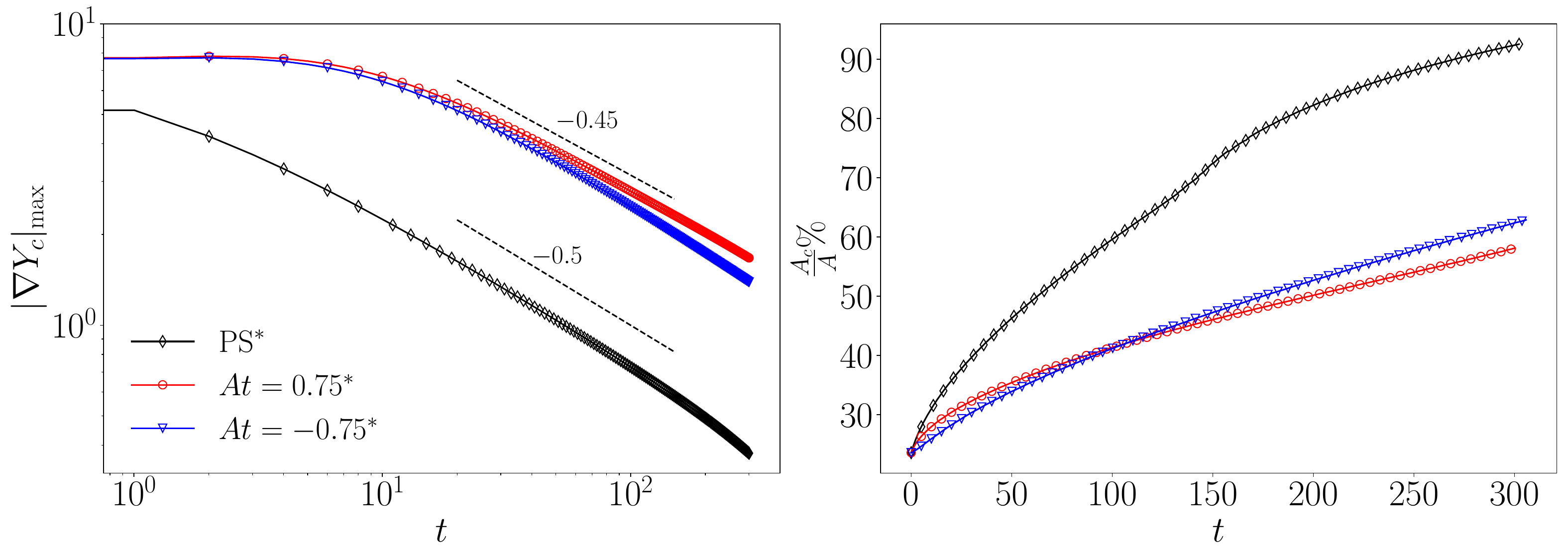}
\put(-375,140){(a)}
\put(-185,140){(b)}
\caption{Evolution of maximum value of $|\bnabla Y_{c}|$ in log-log scale with the averaged slope of decay (a) and evolution of the percentage of area occupied by the circular species (b). The presence of density gradients in active scalars alters the diffusion coefficients as shown in \eqref{eq: Pure Diffusion equation expanded}  compared to passive scalar diffusion, hence modifying the rate of decay. }
\label{fig:Max value of nabla YC}
\end{figure}
\begin{figure}
\includegraphics[width=1.0\textwidth]{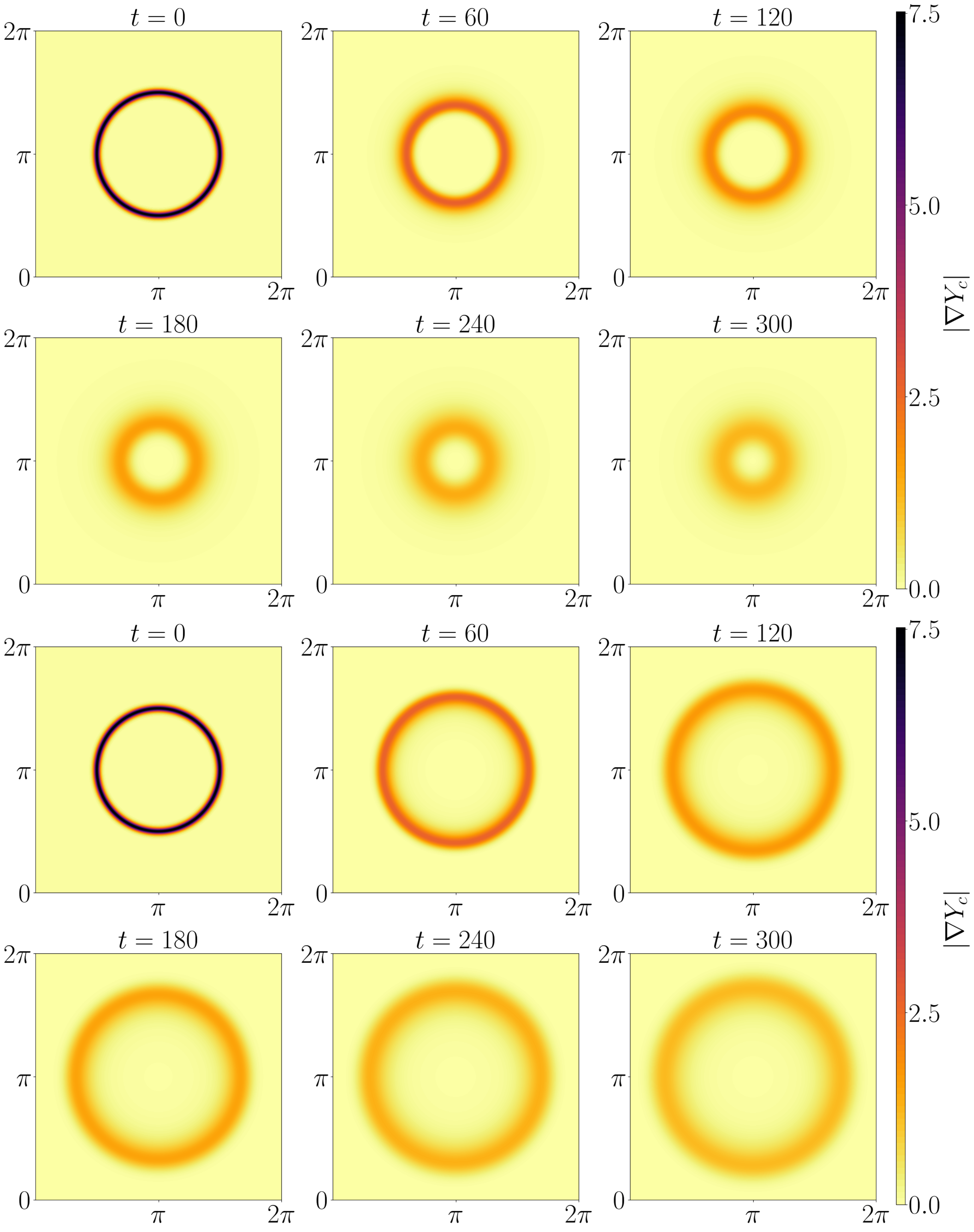}
\put(-400,480){$(a)$}
\put(-400,235){$(b)$}
\caption{Contours of $|\bnabla Y_{c}|$ for $At = 0.75^{*}$ (a) and $At = -0.75^{*}$ (b) at same times. Due to the terms $II$ and $III$ in \eqref{eq: Pure Diffusion equation expanded}, the interface moves inwards for the lighter blob while the interface moves outwards for the heavier blob.}
\label{fig:No shock nabla Y contours}
\end{figure}

Note that for uniform density ($\bnabla \rho =0 , \chi = 1$), \eqref{eq: Pure Diffusion equation expanded} reduces to the standard \Revision{}{linear} diffusion equation of a passive scalar. The parameter $\chi$, which specifies the density ratio between the surrounding and blob species, also indicates the direction of the density gradients. A denser blob has $\chi < 1$ ($D_2 > 0$), while a lighter blob has $\chi > 1$ ($D_2 < 0$). The direction of the density gradient alters the coefficients in the RHS of \eqref{eq: Pure Diffusion equation expanded}. Equation \eqref{eq: Pure Diffusion equation expanded} also indicates that $\bnabla Y_{c}$ is a suitable parameter for quantifying mixing characteristics, and we show in the next section that shock waves increase the mean value of $ |\bnabla Y_{c}|$. Figures \ref{fig:No shock conc contours}(a)  and \ref{fig:No shock conc contours}(b) show the contours of the circular species at different times. The changes in concentration indicate mixing, which in this case occurs solely by molecular diffusion. We choose $|\bnabla Y_{c}|$ to identify the interface between the species, acknowledging that other parameters like mixedness may also be used~\citep{kumar2005stretching}. Figure \ref{fig:Max value of nabla YC}(a)  shows the evolution of the maximum value of $|\bnabla Y_{c}|$, and indicates that the decay of maximum value of $|\bnabla Y_{c}|$ for active scalars is slightly slower compared to the passive scalar. The first term in the RHS of \eqref{eq: Pure Diffusion equation expanded} is the Laplacian term responsible for smoothening out of concentration gradients leading to mixing. In passive scalars, the decay in concentration gradients is solely due to the Laplacian term. However, in active scalars, additional terms ($II$ and $III$ in~\eqref{eq: Pure Diffusion equation expanded}) modify the mixing. Figure \ref{fig:Max value of nabla YC}(b) shows the evolution of the space-filling capacity of the circular blob under the action of diffusion. To quantify the space-filling capacity of the blob species $Y_{c}$, we calculate the area where $Y_{c} > 0.01$ ($A_c$) and compare it to the total area ($A$). The passive scalar under the action of only \Revision{}{linear} Laplacian diffusion process, shows the maximum spread of area, while the nonlinear diffusion process in active scalars limits the spread of the blob. Figure \ref{fig:Max value of nabla YC}(b) also highlights that the movement of the interface of the blob is apparent, and diffusion always spreads the inhomogeneity. The above results indicate that the decay of concentration gradients in active scalars is not as straightforward as passive scalars.

\begin{figure}\centering
\includegraphics[width=1.0\textwidth]{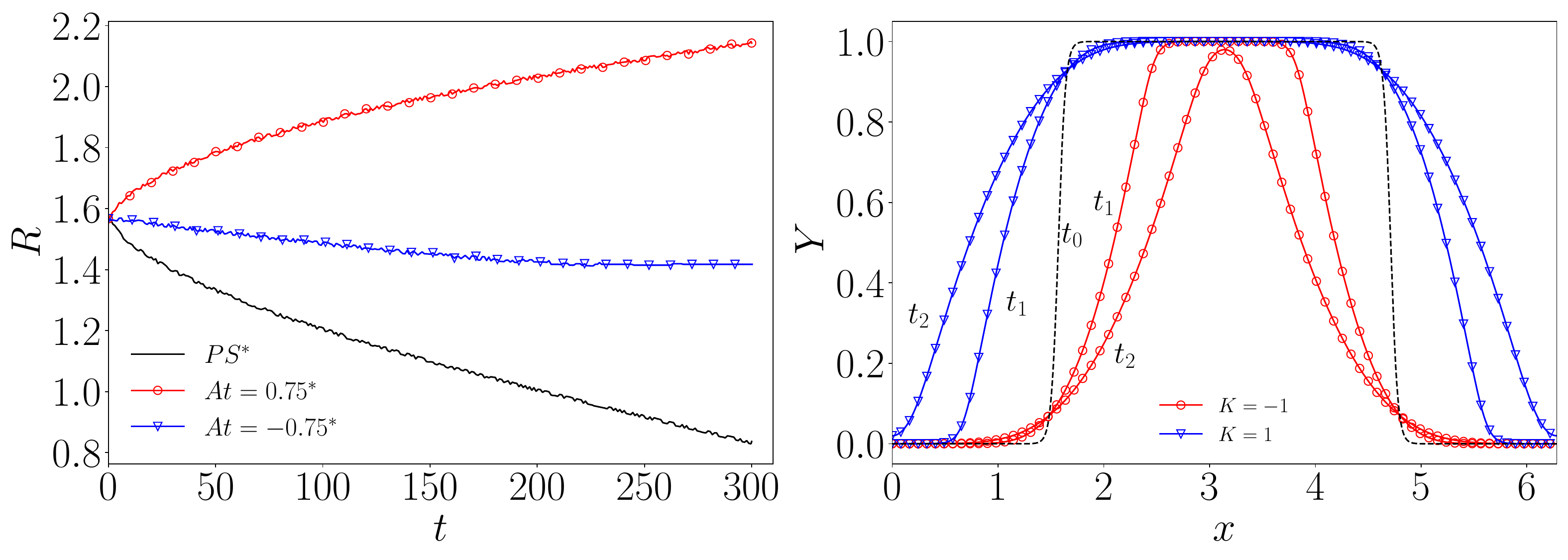}
\put(-375,140){(a)}
\put(-185,140){(b)}
\caption{Evolution of the location of $|\bnabla Y_{c}|_{\mathrm{max}}$ measured from the centre of the blob (a) and one-dimensional evolution of $Y$ \Revision{}{from the 1D unsteady nonlinear diffusion equation} \eqref{eq: 1D dissipation} solved using a 1D pseudospectral solver (b). The directions of concentration and density gradients affect the coefficients of terms $II$ and $III$ in ~\eqref{eq: Pure Diffusion equation expanded}, causing the interface of the positive Atwood case to move inward and the negative Atwood case to move outward. The black line represents the initial condition at $t = t_{0}$. The blue and red lines represent cases \Revision{}{with $K=\pm 1$, respectively and $t_{0} < t_{1} < t_{2}$. For $K>0$, the interface moves outward while diffusing and for $K<0$, the interface moves inward while diffusing. }}

\label{fig:Location of gradY and 1D setup}
\end{figure}

The second and third terms in \eqref{eq: Pure Diffusion equation expanded} are additional terms modifying the diffusion of active scalars. The effect of these terms is shown in the contours of $|\bnabla Y_{c}|$ at different time instances in figures \ref{fig:No shock nabla Y contours}(a) and \ref{fig:No shock nabla Y contours}(b), for $At>0$ and $At<0$ respectively. We see that for the $At > 0$ (denser mixture, lighter blob), the interface moves inward, and the blob seems to shrink, while for $At < 0$ (lighter mixture, denser blob), the interface moves outward, and the blob seemingly expands. This highlights the biased diffusion process in active scalars, where the interface moves towards the lighter species. The two terms mentioned above ($II$ and $III$ in \eqref{eq: Pure Diffusion equation expanded}) are similar to the scalar dissipation term analysed in the mixing of shock-bubble interaction studies and hence, are referred to as \emph{\Revision{}{nonlinear} dissipation terms} \citep{tomkins2008experimental,yu2021scaling}. Specifically, we choose to call $II$ and $III$ in \eqref{eq: Pure Diffusion equation expanded} as concentration gradient driven and density gradient driven dissipation, respectively, in the current study. Figure \ref{fig:Location of gradY and 1D setup}(a) shows the evolution of radius $R$ at which the maximum value of $ |\bnabla Y_{c}|$ is located from the centre of the blob. We see that for the lighter blob case ($At = 0.75$), the interface moves inward towards the centre, while for the heavier blob case ($At = -0.75$), the interface moves away from the centre. We also see that in the passive scalar diffusion, there is very little movement of the interface. To illustrate that the \Revision{}{nonlinear} dissipation terms ($II$ and $III$ in \eqref{eq: Pure Diffusion equation expanded}) are indeed responsible for the movement of the interface, we solve a 1D \Revision{}{unsteady nonlinear conduction equation,
 \begin{equation}
     \frac{\partial Y}{\partial t} = D\frac{\partial^2 Y}{\partial x^2}+K \left(\frac{\partial Y}{\partial x} \right)^{2},
     \label{eq: 1D dissipation}
 \end{equation}
 with $K=\pm 1$ for the same initial condition as \eqref{eq: circular species ditribution} and $D=1/10$ in 1D.
}
 
  \Revision{}{Figure \ref{fig:Location of gradY and 1D setup}(b) shows the evolution of $Y$ governed by \eqref{eq: 1D dissipation} with $K=\pm 1$. We see that for positive coefficients, the interface moves outward, while for negative coefficients, the interface moves inward. Hence, the essential physics of the movement of the interface is captured by the nonlinear conduction equation~\eqref{eq: 1D dissipation} highlighting that the nonlinear diffusion terms in \eqref{eq: Pure Diffusion equation expanded} result in the movement of the interface, while the Laplacian term diffuses the interface as usual. These observations agree with the known diffusive travelling wave type solutions of nonlinear conduction equations~\citep{atkinson1981traveling}. Additionally,} these results are also in line with the observations in figure ~\ref{fig:Location of gradY and 1D setup}(a), where from \eqref{eq: Pure Diffusion equation expanded}, we see that for $At = 0.75,~\chi > 1,~D_2 < 0$ and hence the interface moves inward, while for $At = -0.75,~\chi < 1,~D_2 > 0$ and hence the interface moves outward. Analogously, the density driven nonlinear dissipation term, alters the direction of movement of the interface based on the density gradient direction. The movement of the interface affects the decay of concentration gradients and hence we see the difference in average slopes of decay in figure \ref{fig:Max value of nabla YC}(a).

 Although the interface (location of $ | \bnabla Y_{c}|_{\mathrm{max}}$) seems to expand for $At < 0 $ and shrink for $At > 0$, the area \Revision{}{where the blob species can be found (or the area occupied) identified as $Y_c>0.01$ increases for both the cases.} Figure \ref{fig:Max value of nabla YC}(b) shows that the area occupied is similar for both $At < 0$ and $At > 0$, \Revision{}{but significantly smaller than the passive scalar. In the next section, we discuss the influence of shock waves on active scalar mixing}. 

\end{subsection}
\begin{subsection}{ Stirring effect of shock waves}
\label{sec:Stirring effect of shock waves}
\begin{figure}
 	\includegraphics[width=\textwidth]{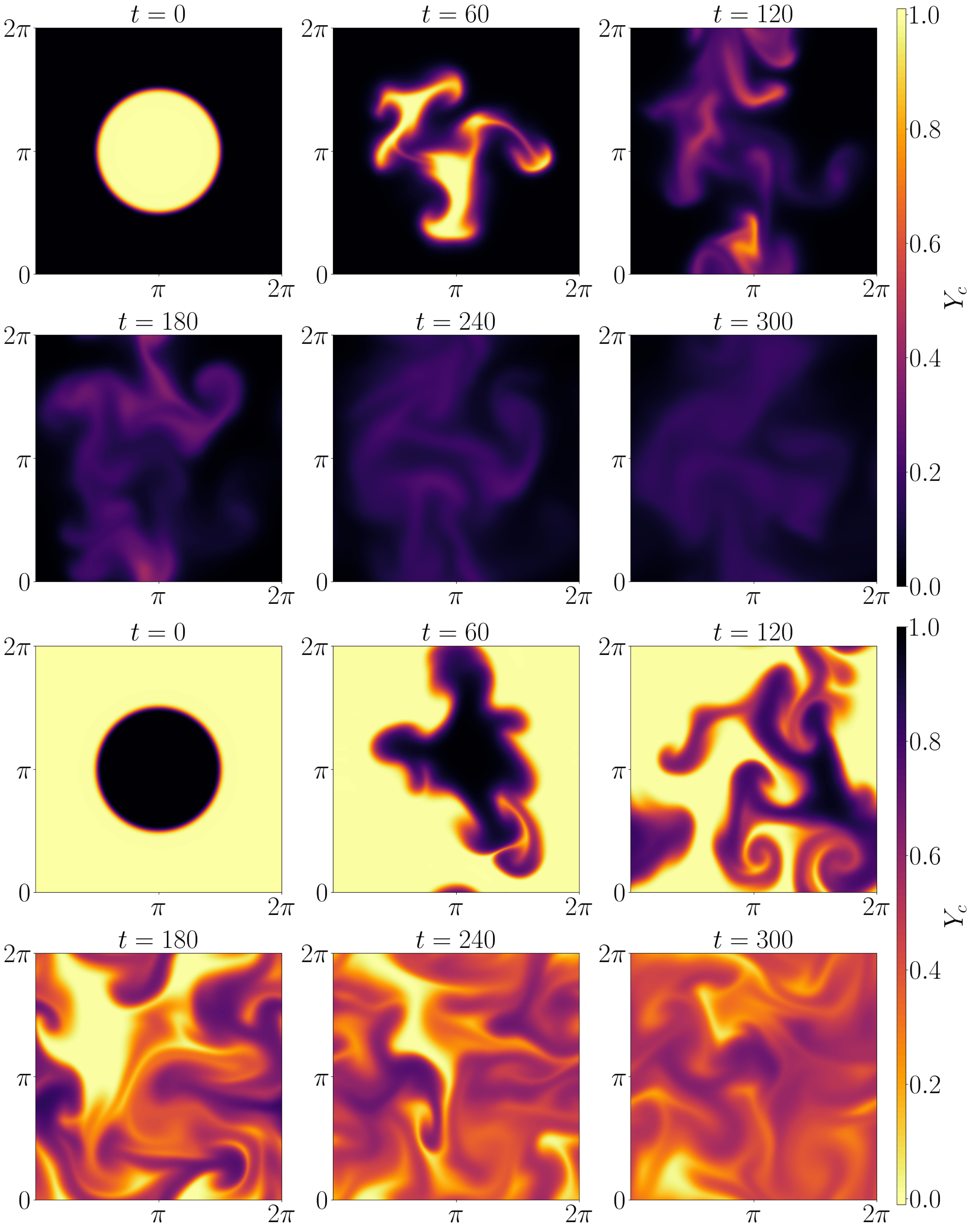}
 	\put(-400,480){$(a)$}
 	\put(-400,235){$(b)$}
 	\caption{Contours of concentration of blob for $At = 0.75$ (a) and $At = -0.75$ (b) at same times. Darker colour indicates heavier species. A comparison with figures \ref{fig:No shock conc contours}(a) and \ref{fig:No shock conc contours}(b) shows that shock waves enhance the mixing process. Shock waves continuously deform the interface of the blob, breaking it apart. The denser mixture (lighter blob) homogenizes faster than the lighter mixture (denser blob). In the above contours, the lighter colour indicates the species that is less dense while the darker colour indicates the species that is heavier.}
  \label{fig:Shock conc contours}
 \end{figure}

Stirring increases the mean value of concentration gradients and also increases the area across which molecular diffusion occurs. Random shock waves interact with the blob interface characterized by large density gradients. Baroclinic torque generated by such interactions generates localized vorticity at the interface. Consequently, the interface deforms and folds, thus increasing the apparent perimeter over which diffusion takes place. We call this baroclinic vorticity induced mixing as stirring. In this section, we analyse the stirring action of shock waves and its role in enhancing the mixing of active scalars. We also discuss the effect of the Atwood number on the evolution of the concentration levels and present a comparison of the time scales of each mixing process. Furthermore, we distinguish between stirring-dominated and diffusion-dominated regimes of mixing by using the maximum value of species concentration.  
 
Stochastically generated shock waves interact with the blob interface where the gradient of density is maximum. Figures \ref{fig:Shock conc contours}(a) and \ref{fig:Shock conc contours}(b) show the contours of the evolution of the circular species concentration for $At = 0.75$ and $At = -0.75$, respectively. In comparison with the no shock cases of figures \ref{fig:No shock conc contours}(a)  and \ref{fig:No shock conc contours}(b), we see that in the presence of shock waves, the blob is stretched and folded leading to a continuous distortion of the interface. The figures suggest that homogenisation occurs at a faster rate under the stirring action of shock waves, resulting 
 in a completely mixed state. To quantify the extent to which shocks enhance the mixing process, we compare the space-filling capacity of the blob species in the presence of shock waves to the no-shock cases. Figure \ref{fig:Mixing percentage and grad Y}(a) shows that shock waves increase the rate of the space-filling capacity of the blob compared to the no-shock/only molecular diffusion case. Shock waves increase the space-filling capacity by increasing the mean concentration gradients and sustaining them for a longer time as shown in figure \ref{fig:Mixing percentage and grad Y}(b). \Revision{}{For passive scalars, shocks have negligible affect on the mixing because the baroclinic torque is absent.} We discuss the reason for the sustained gradients \Revision{}{in the case of active scalars} in detail below.

 \begin{figure}\centering
\includegraphics[width=1.0\textwidth]{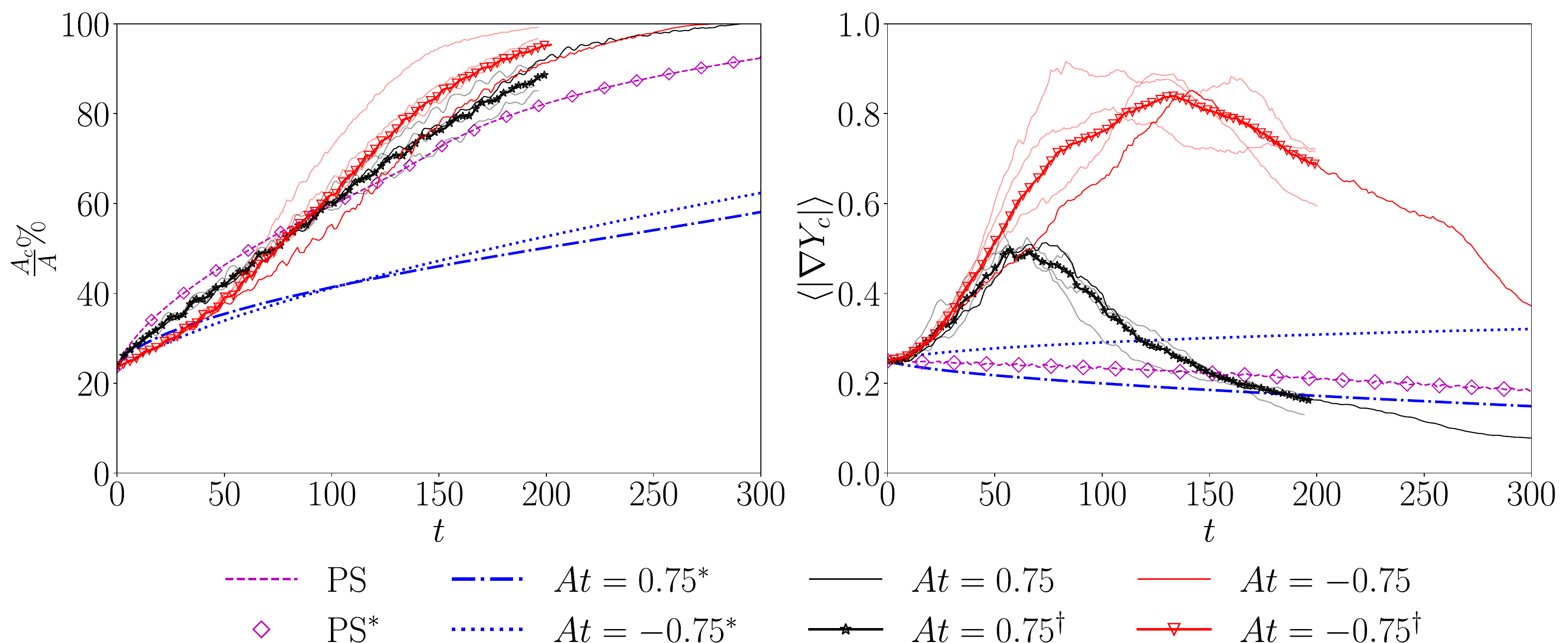}
\put(-375,160){(a)}
\put(-185,160){(b)}
\caption{\Revision{}{Evolution of percentage of the ratio of mixed area to total area (a) and the evolution of area-averaged concentration gradients (b). Shock waves improve the space-filling capacity of active scalars when compared to the pure diffusion cases. The labels are identical for both the figures.}}
\label{fig:Mixing percentage and grad Y}
\end{figure}

\begin{figure}
\includegraphics[width=1.0\textwidth]{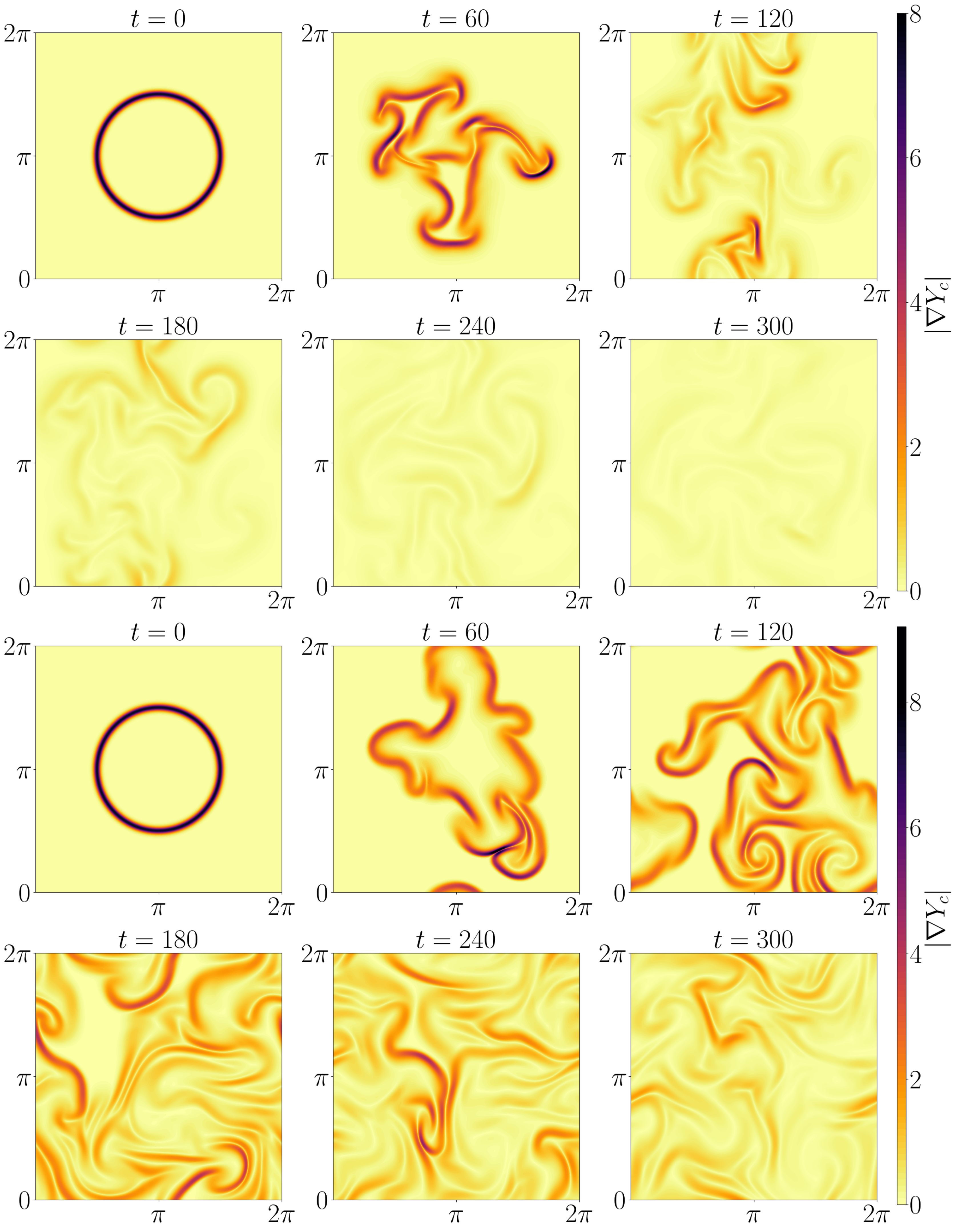}
	\put(-400,480){$(a)$}
	\put(-400,235){$(b)$}
	\caption{Contours of $|\bnabla Y_{c}|$ for $At = 0.75$ (a) and $At = -0.75$ (b) at same times. The concentration gradients are sustained longer for the denser blob. This is due to the outward expansion of interface by the terms $II$ and $III$ in \eqref{eq: Pure Diffusion equation expanded} which delays the exposure of the blob to the action of molecular diffusion.}
	\label{fig:shock nabla Y contours}
\end{figure}

 The concentration diffusion flux is proportional to the perimeter of the interface across which diffusion occurs and the strength of the gradients \Revision{}{across the interface}. As mentioned earlier, the role of stirring is to increase the perimeter, thereby increasing the total diffusion flux across the interface. This also results in an increase in the mean magnitude of concentration gradients. Figures \ref{fig:shock nabla Y contours}(a) and \ref{fig:shock nabla Y contours}(b) show the contours of $|\bnabla Y_{c}|$  for positive and negative Atwood numbers, respectively. The figures also show that the mean magnitude of concentration gradients ($\langle |\bnabla Y_{c}| \rangle$) increases due to an increase in the perimeter of the blob interface caused by shock waves. Furthermore, for all the active scalar cases with shocks, we see that the concentration gradients decay faster in the final time instances when compared to the no-shock cases. Figure \ref{fig:shock nabla Y contours}(a) shows that a lighter blob homogenizes faster when compared to a denser blob. To quantify the effectiveness of stirring, we look at the mean magnitude of concentration gradient and compare it with the mixing driven by only molecular diffusion in figure~\ref{fig:Mixing percentage and grad Y}(b). For the no-shock cases, the negative Atwood number case has a mean value that increases with time, while for the positive Atwood number case, the mean value of the concentration gradient decays with time. This is due to the nonlinear diffusive transport (termed as \Revision{}{nonlinear} dissipation) which causes the interface to expand for $At < 0$ and shrink for $At > 0$. Initially, when $Y_c \approx 1$ near the interface, the denser blob  ($At <0 $) tends to expand quickly compared to diffusion, and vice versa for the lighter blob ($At > 0$). This can be realised by considering the ratio of the molecular diffusion time scale ($ \tau_{m}$) to the \Revision{}{nonlinear} dissipation time scale ($ \tau_{d}$) given as
 \begin{equation}
     \frac{\tau_{m}}{\tau_{d}} \sim \frac{|D_{2}|}{D_{1}} \sim \frac{|1 - \chi|}{1 + \left( \chi -1 \right) Y_{c}}.
     \label{eq:time scale of molecular to dissipation}
 \end{equation}
The reader is referred to Appendix \ref{appB} for a detailed discussion on the ratio of the molecular diffusion time scale to the \Revision{}{nonlinear} dissipation time scale. 
 \begin{figure}\centering
\includegraphics[width=1.0\textwidth]{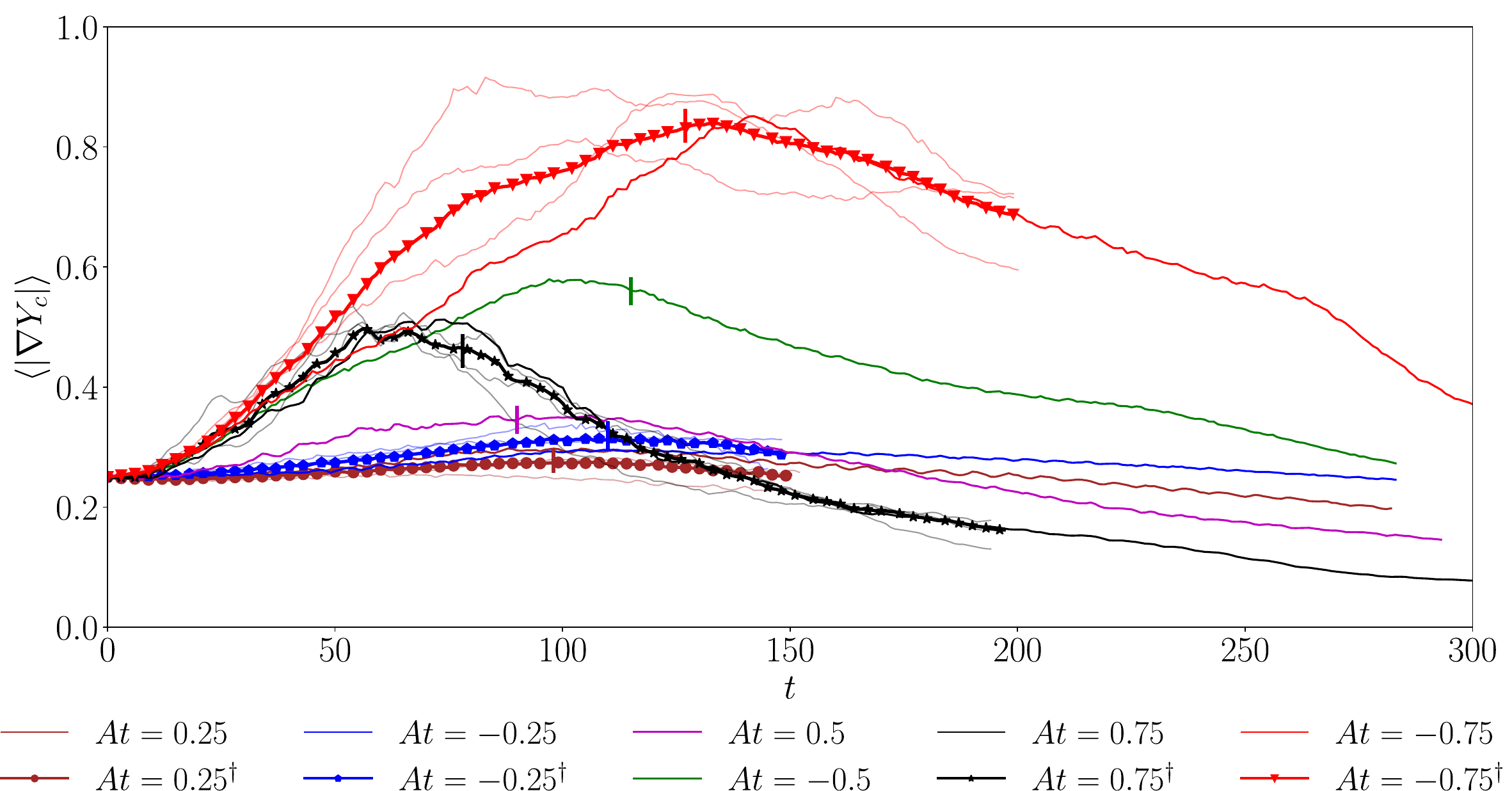}
%\put(-375,140){(a)}
%\put(-185,140){(b)}
\caption{\Revision{}{The evolution of area-averaged concentration gradients. The vertical lines indicate the mixing time calculated using Eq \eqref{eq: Normalised max conc} and \eqref{eq: Normalised max conc for negative Atwood numbers} for positive and negative Atwood numbers, respectively. The labels are identical for both the figures.}}
\label{fig: Grad Y_c of Atwood numbers}
\end{figure}

Figure \ref{fig:Mixing percentage and grad Y}(b) shows that the shock waves increase the mean magnitude of concentration gradients for the  Atwood numbers cases. However, passive scalars concentration gradient is not affected by the shock waves. Figure \ref{fig: Grad Y_c of Atwood numbers} shows that shock waves increase the mean magnitude of the concentration gradients for all Atwood numbers considered by increasing the area (apparent perimeter in 2D) across which the gradients are present. Figure \ref{fig: Grad Y_c of Atwood numbers} also shows that the negative Atwood numbers attain a higher value of $\langle |\bnabla Y_c| \rangle$ than the positive Atwood numbers. Though the diffusion term smoothens out the gradients and reduces the value of the concentration gradients in all Atwood number cases locally, the increase in the mean magnitude is due to the increase in the perimeter. The outward expansion in the negative Atwood number cases further increases the perimeter and causes the gradients to exist over a larger perimeter. \Revision{}{Additionally, as the magnitude of the Atwood number increases, more stretching and folding of the interface occurs thus increasing the generated maximum mean concentration gradients.}

Figure \ref{fig: Max nabla Yc with shock cases} shows the evolution of maximum value of $|\bnabla Y_c|$ for passive and active scalar cases with shocks. We see that the \Revision{}{dynamics of the }passive scalar case with \Revision{}{and without} shocks are similar. A few fluctuations in the maximum value of $|\bnabla Y_c|$ exist which can be attributed to the randomized vorticity generated due to the curvature of the randomly generated shocks~\citep{jossy2023baroclinic}. The maximum value of $|\bnabla Y_c|$ also decays for the active scalar cases but with localised random spikes in values. The presence of shock waves sustains the gradients longer for the active scalar cases. We quantify this through figure \ref{fig: PDF Species diffusion gradient time series}(a) and \ref{fig: PDF Species diffusion gradient time series}(b), which show the probability density function (PDF) of the magnitude of concentration gradients for $At=0.75$ and $At=-0.75$, respectively. We observe that the decay of the tail is more rapid for $At > 0$ compared to the case of $At < 0$. 

The above discussion allows us to distinguish two regimes of mixing -- one where the action of stirring is dominant (termed as the vorticity--dominated regime) and the other where molecular diffusion is dominant (termed as the diffusive regime). The regime to the left of the peak value of $\langle |\bnabla Y_{c}| \rangle$ in figure \ref{fig: Grad Y_c of Atwood numbers} is the vorticity---dominated regime, and the regime to the right is the diffusive regime. In the vorticity--dominated regime, the interface gets stretched and folded due to baroclinic vorticity (also discussed in \S\ref{sec: baroclinicity}). As the stretching continues, more points inside the blob are exposed to the surrounding species, thus facilitating molecular diffusion more. As we discuss below, for $At<0$, the interface keeps expanding, thus delaying the time taken by stirring to expose all the points inside the blob to the surrounding species.
% The stretching of the interface continues till all points of the blob are exposed to the surrounding species.
\begin{figure}\centering
\includegraphics[width=1.0\textwidth]{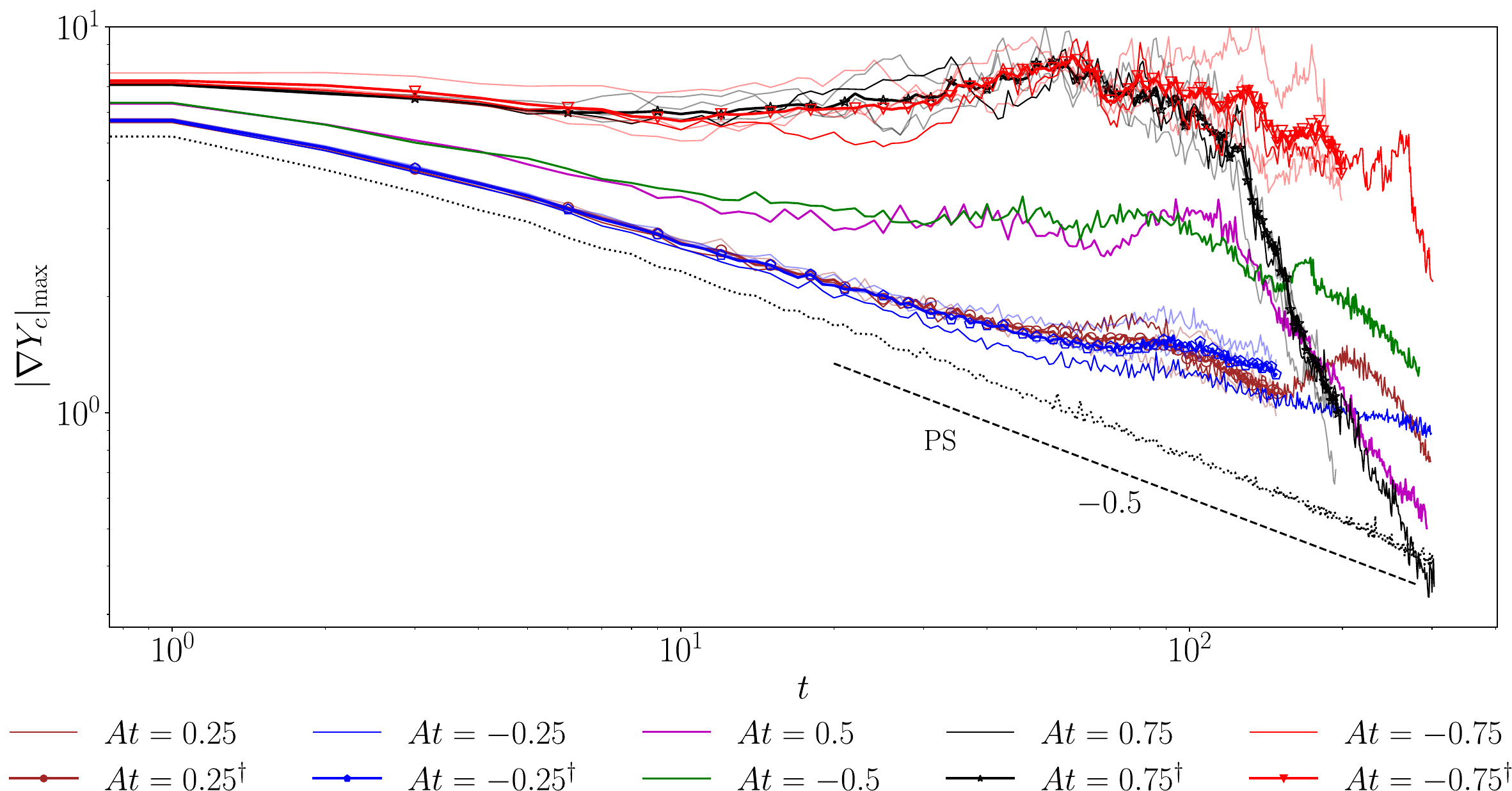}
	\caption{\Revision{}{Evolution of maximum value of $|\bnabla Y_{c}|$ in log-log scale for all the cases in table \ref{tab: shock wave parameter space}. Shock waves have no significant effect on the decay of the maximum concentration gradients of passive scalars. The presence of density gradients in active scalars sustains concentration gradients longer compared to the passive scalar.}}
	\label{fig: Max nabla Yc with shock cases}
\end{figure}

To estimate the time till which the interface of the blob can be stretched, we calculate the normalised maximum concentration ($Y^{N}_{c}$) given by,
\begin{equation}
    Y^{N}_{c} = \frac{Y_{c,\mathrm{max}}}{Y^{0}_{c,\mathrm{max}}},
    \label{eq: Normalised max conc}
\end{equation}
where $Y_{c,\mathrm{max}}$ is the maximum concentration of the circular species at the given instant and $Y^{0}_{c,\mathrm{max}}$ is the initial maximum concentration of the circular species. The normalised maximum concentration has been used to estimate the mixing time in passive scalar mixing~\citep{meunier2003vortices} and shock-bubble interaction mixing~\citep{liu2022mixing} studies. A normalised maximum concentration ($Y^{N}_{c}$) $< 1$ indicates that stirring has stretched the blob interface and exposed all of it to the surrounding species.  After this instant, the role of stirring is secondary and the molecular diffusion dominates. For positive Atwood numbers, the \Revision{}{nonlinear} dissipation results in the apparent contraction of the blob. Hence, the overall time taken to expose all the points of the blob species to the surroundings is smaller. However, for negative Atwood numbers, the effect of the outward expansion delays the time taken for all points of the blob to be exposed to the surrounding species. To account for the outward expansion, we modify the normalised maximum concentration as follows for negative Atwood number cases,
\begin{equation}
    Y^{N}_{c} = \frac{Y_{c,\mathrm{max}}}{ 0.95 \hspace{1.5mm} Y^{0}_{c,\mathrm{max}
    }}.
    \label{eq: Normalised max conc for negative Atwood numbers}
\end{equation}

The vertical lines on figure \ref{fig: Grad Y_c of Atwood numbers} indicate the time when normalised maximum concentration falls below 1. We see that the estimated time is located very close to the peaks of $\langle |\bnabla Y_{c}| \rangle$. Hence, the calculated normalised maximum concentration highlights that stirring deforms any geometrical concentration inhomogeneity till all the points of the inhomogeneity are exposed to stronger gradients. Therefore, it serves as a suitable indicator for distinguishing the regimes where stirring and molecular diffusion are dominant. In the next section, we discuss the role of baroclinic vorticity and also its role in increasing the stirring time of a denser blob.

\end{subsection}

\begin{subsection}{Role of baroclinicity in stirring }
\label{sec: baroclinicity}
 In this section, we discuss the mechanism by which the stirring action of the randomly generated shock waves increases the apparent interfacial perimeter. We show the effect of the Atwood number on baroclinicity and its role in enhancing the mixing. We also derive a time scale for the shock-induced stirring and show that it is sustained for a longer time for a lighter mixture.

 As discussed in the previous section, the presence of shock waves increases the mean magnitude of concentration gradients for active scalars. But, figure \ref{fig:Mixing percentage and grad Y}(b) shows that in spite of the presence of shocks, the mean magnitude of the concentration gradients for the passive scalar does not increase. The stirring action of shock waves affects mixing through the advection term of the species equation in \eqref{eq: species equation}. The velocity in the advection term can be decomposed using Helmholtz decomposition as follows,
\begin{equation}
    \boldsymbol{u} = \boldsymbol{u_{w}} + \boldsymbol{u_{v}}
    \label{eq: velocity using Helhmholtz decomp}
\end{equation}
where $\boldsymbol{u_{w}}$ and $\boldsymbol{u_{v}}$ are the acoustical and vortical components of velocity, respectively. Both passive and active scalars have the same rate of injection of acoustical energy. Therefore, any difference caused between the two would stem from the vortical component of velocity ($\boldsymbol{u_{v}}$). We obtain the evolution equation of vorticity ($\boldsymbol{\omega}$) by taking the curl of \eqref{eq: momentum_equation} as,
%\begin{equation}
%\frac{\partial  \boldsymbol{\omega} }{\partial t}=
% \frac{\bnabla \rho \times \bnabla p}{\rho^2} +  \frac{\mu\bnabla^{2} \boldsymbol{\omega}}{\rho \mathrm{Re}_{\mathrm{ac}}}  +\frac{\mu}{\mathrm{Re}_{\mathrm{ac}}} \frac{\bnabla \rho \times [ \bnabla \times \boldsymbol{\omega} ] }{\rho^2}-  \left (\frac{4 \mu}{3\mathrm{Re}_{\mathrm{ac}}} \right) \frac{\bnabla \rho \times \bnabla (\bnabla \bcdot \boldsymbol{u} ) }{\rho^{2}}  -\bnabla\bcdot(\boldsymbol{u}\boldsymbol{\omega}) + \boldsymbol{F^r_\omega} .
%\label{eq: omega}    
%\end{equation}

\begin{figure}\centering
\includegraphics[width=1.0\textwidth]{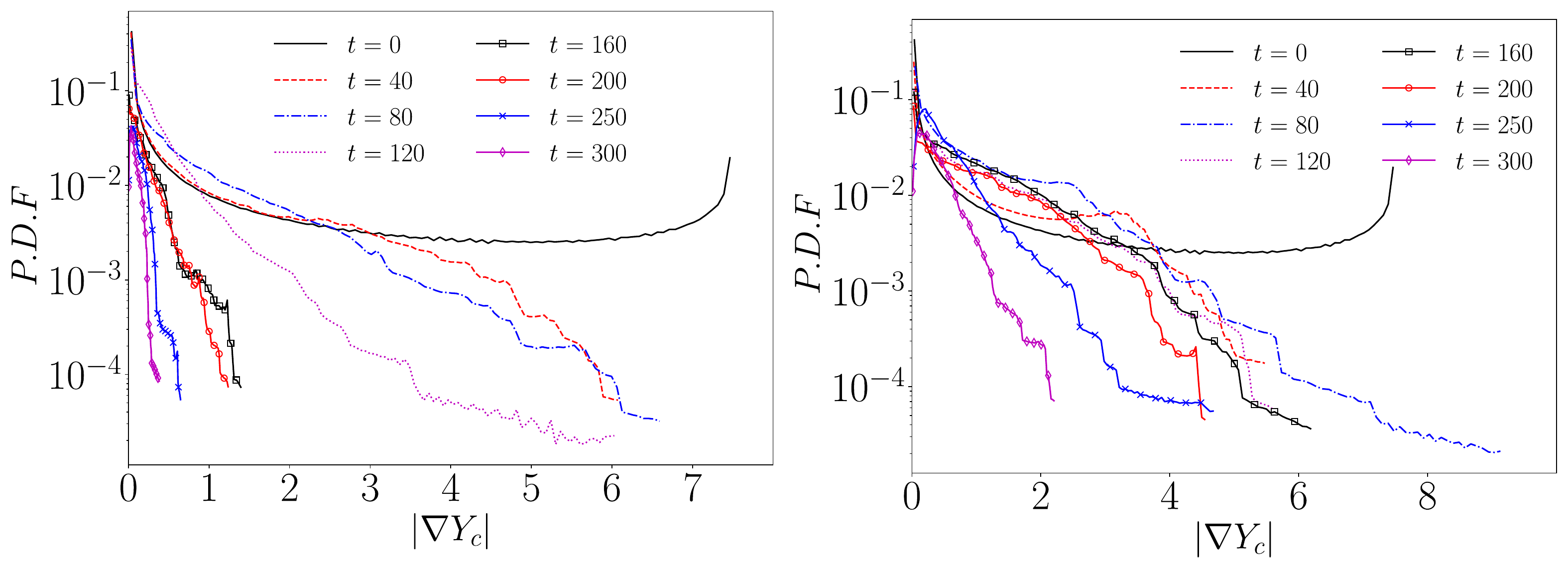}
\put(-375,140){(a)}
\put(-185,140){(b)}
\caption{PDF of the magnitude of concentration gradient at different time instances for $ At = 0.75$ (a) and $ At = -0.75$ (b). Initially, the PDF indicates two peaks, one at 0 and other at the maximum value inside the interface. With time, the PDF tails drop indicating diffusion. Negative Atwood numbers sustain concentration gradients longer owing to the outward expansion effect of the \Revision{}{nonlinear} dissipation terms.}
\label{fig: PDF Species diffusion gradient time series}
\end{figure}

\begin{eqnarray}
\frac{\partial  \boldsymbol{\omega} }{\partial t}  & = &
 \frac{\bnabla \rho \times \bnabla p}{\rho^2} +  \frac{\mu\bnabla^{2} \boldsymbol{\omega}}{\rho \mathrm{Re}_{\mathrm{ac}}}  +\frac{\mu}{\mathrm{Re}_{\mathrm{ac}}} \frac{\bnabla \rho \times [ \bnabla \times \boldsymbol{\omega} ] }{\rho^2} \nonumber \\
&& -  \left (\frac{4 \mu}{3\mathrm{Re}_{\mathrm{ac}}} \right) \frac{\bnabla \rho \times \bnabla (\bnabla \bcdot \boldsymbol{u} ) }{\rho^{2}}  -\bnabla\bcdot(\boldsymbol{u}\boldsymbol{\omega}) + \boldsymbol{F^r_\omega} .
\label{eq: omega}
\end{eqnarray}
%\begin{figure}\centering
%\includegraphics[width=0.7\textwidth]{./Figures/Passive_Scalar_nablaYc.pdf}% Here is how to import EPS art
%\caption{ Evolution of maximum value of $|\bnabla Y_{c}|$ for passive scalar with and without shock waves. Shock waves do not enhance mixing by stirring in the absence of vorticity. }
%\label{fig: Passive case nabla Yc}
%\end{figure}

The last term in the RHS of \eqref{eq: omega} is the vortical forcing component and is zero in the current study. The first term in the RHS is the baroclinic term and is the primary term responsible for generation of vorticity in the domain, and hence the stirring action.  \Revision{reviewer 1}{ The behavior of the diffusion terms of vorticity is similar to the findings reported earlier~\citep{jossy2023baroclinic}, where the interaction of random shock waves with thermal gradients was analyzed. The second term is the
	viscous diffusion of vorticity and is the major  contributor to the vorticity dissipation. The next two terms represent the interaction of density gradients with viscous stresses in the domain. The term proportional to bulk force ($\bnabla(\bnabla \bcdot \boldsymbol{u})$) has higher dissipation contribution than the dissipation term propositional to the shear stress ($\bnabla \times \boldsymbol{\omega}$) in a flow field where shocks are involved. The term proportional to bulk force will be maximum along the shocks where the divergence of velocity is maximum. These terms will increase in magnitude with increase in Atwood number, due to stronger density gradients. The second last term is the advection term of vorticity, which has the  minimum contribution to the enstrophy budgets~\citep{jossy2023baroclinic}. }
 
 \Revision{1}{The difference between passive and active scalars is the presence of the density gradients in the latter. In our study, vorticity production is primarily due to the misalignment of the pressure gradients in the shock waves and the density gradients at the blob interface. This indicates that the absence of strong density gradients in the passive scalar case results in no stirring effect. The mixing of both active and passive scalars is primarily due to the vortical (solenoidal) structures and is independent of compressibility when it comes to large-scale effects.}  Similar observations were made in the compressible mixing of passive scalars, where the dilational component did not enhance the mixing process of passive scalars \citep{john2019solenoidal,ni2016compressible}. \citet{jossy2023baroclinic} show that shock waves generated by stochastic forcing generate broadband vorticity owing to the continuous variation of shock curvature. We see that vorticity generated due to shock curvature has only a slight and insignificant effect on the mixing of passive scalars. Hence, the vorticity generated at the interface between the blob and surrounding species is solely responsible for mixing of the active scalars.
\begin{figure}
\includegraphics[width=1.05\textwidth]{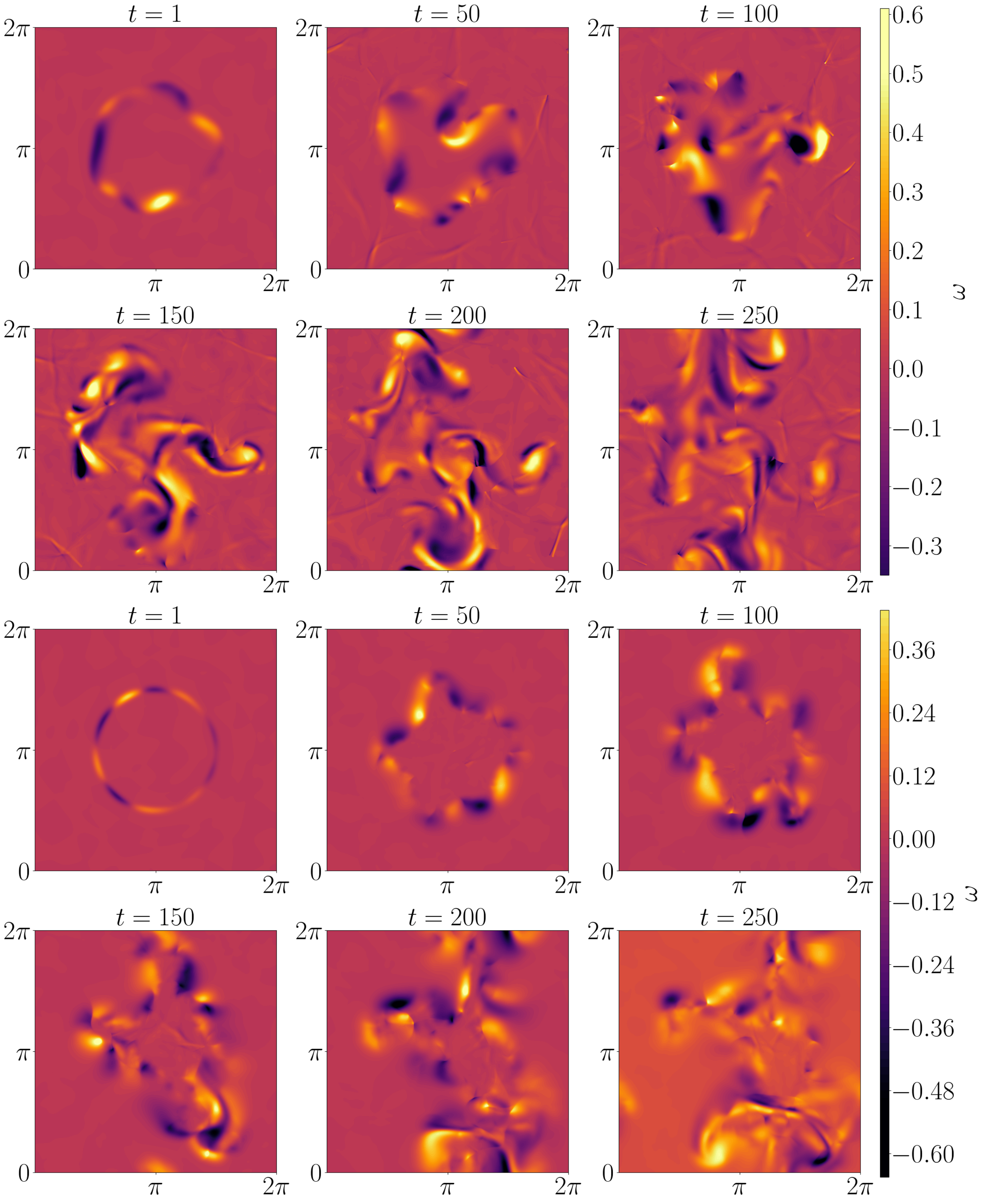}
\put(-420,480){$(a)$}
\put(-420,235){$(b)$}
\caption{Contours of vorticity for $At = 0.75$ (a) and $At = -0.75$ (b) at same times. The vorticity is concentrated along the perimeter where density gradients are maximum with alternating rotating and counter-rotating vortices. Spikes and bubbles develop at the intersection of the vortices rotating in opposite directions.}
\label{fig: Vorticity Contours At=-0.75}
\end{figure}

Figures \ref{fig: Vorticity Contours At=-0.75}(a) and \ref{fig: Vorticity Contours At=-0.75}(b) show the vorticity contours of the $At= 0.75$ and $At= -0.75$, respectively. We see that the vorticity is concentrated along the interface of the blob. As discussed earlier, the only source of production of strong vorticity in a two-dimensional compressible flow in a periodic domain is baroclinicity. The pressure gradients within the shock waves and the high-density gradients along the interface of the inhomogeneity result in vorticity production due to baroclinicity. In this context, the current study resembles a continuously forced RMI setup \citep{meshkov1969instability,richtmyer1954taylor}, where shock waves are continuously distorting the interface at random locations and random time intervals governed by the stochastic forcing of the shock waves. Similar to RMI, we see the development of spikes (denser fluid penetrating into lighter fluid) and bubbles (lighter fluid penetrating into denser fluid) at the interface \citep{abarzhi2002new,yuan2023instability}. Figures \ref{fig: Vorticity Contours At=-0.75}(a) and \ref{fig: Vorticity Contours At=-0.75}(b) show that both clockwise and counter-clockwise rotating vortices are present along the interface of the blob and surrounding species. The resultant direction of velocity at each intersection of clockwise rotating vortices and counter-clockwise rotating vortices results in either the production of a spike or a bubble. This results in the stretching of the interface in addition to the stretching (compression) occurring from the nonlinear dissipation terms for a denser (lighter) blob (as highlighted in \S\ref{sec:Effect of density gradients on molecular diffusion}). The sharp jump in tangential velocity across the interface can also trigger shear instability, a Kelvin-Helmholtz instability (KHI) in the limiting inviscid case. Such shear instability results in further folding of the interface. However, in the current study, it is difficult to distinguish between the role of  RMI and KHI in enhancing the mixing due to the continuous forcing of the RMI. We also see a reduction in the vorticity production with time. This is expected as the mixing in active scalars proceeds to homogenise density gradients, thus reducing the magnitude of baroclinic torque generated due to the interaction of shock waves. \Revision{1,2}{Note that the shock waves are continuously forced, and hence typical values of $\bnabla p$ remain similar as indicated by Figure \ref{fig: Mach contour and Mach PDF}(b). The strength of the shock waves can be altered by changing the rate of energy injection $\varepsilon$ in \eqref{eq: forcing_injection}. Any reduction  in the rate of energy injection will reduce the strength of the shock waves, leading to a weaker baroclinic vorticity. This will result in the species taking a longer time to homogenise. In a similar fashion, any increase in rate of energy injection will result in stronger shocks leading to faster homogenisation. } 

\Revision{}{At the initialization, the magnitude of the density gradient increases with the absolute value of the Atwood number due to the increasing density difference.  Stronger density gradients produce higher values of baroclinic vorticity. Figure \ref{fig: Area averaged baroclinic vorticity}(a) shows the evolution of the magnitude of the  area-averaged baroclinic vorticity generation for all the active scalar cases with shock waves. We see that the baroclinicity increases with a wider disparity in density ratio between the surrounding and the circular species. The increase in baroclinicity indicates an increasing amount of stirring}. Furthermore, prolonged high values of baroclinicity indicate prolonged stirring, which is the case for $At<0$ mixtures. Figure~\ref{fig: Area averaged baroclinic vorticity}(a) suggests that baroclinicity is responsible for stretching the interface and that the stretching is higher for larger absolute values of the  Atwood number. To show that the stretching of the interface is primarily due to stirring and not the  nonlinear dissipation terms, we compare their respective time scales. We derive the time scale for stirring from the evolution equation for the advection term of the species equation, which is given by 
 \begin{equation}
    \frac{\partial Y_{k}}{\partial t} = \boldsymbol{u_v} \cdot \bnabla Y_{k}.
\end{equation}
\begin{figure}\centering
\includegraphics[width=1.0\textwidth]{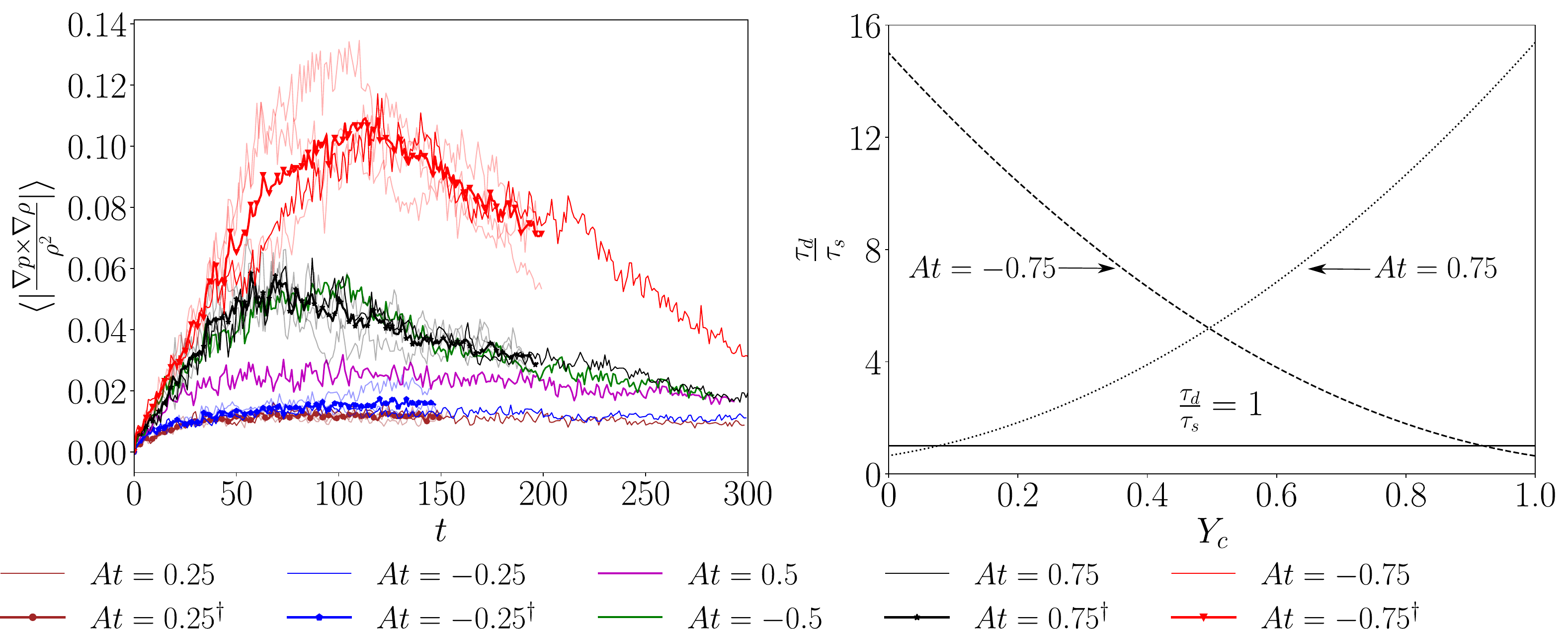}% Here is how to import EPS art
\put(-375,160){(a)}
\put(-185,160){(b)}
\caption{ \Revision{}{Evolution of the area-averaged magnitude of baroclinic vorticity generation (a) and the variation of the ratio of the time scale of stirring to nonlinear dissipation with a concentration level of blob (b). The wider disparity in density ratios results in stronger baroclinic vorticity production. The stronger vorticity is responsible for stretching the perimeter of the interface.}}
\label{fig: Area averaged baroclinic vorticity}
\end{figure}
Using scaling arguments, the time scale of stirring $\tau_{s}$ can be written as 
\begin{equation}
    \frac{1}{\tau_{s}} \sim \left(\frac{L_{v}}{\tau_{v}}\right) \frac{1}{L},
    \label{eq: time scale of stirring result }
\end{equation}
where $\tau_{v}$ is the vortical time scale, $L$ is the length scale across which the concentration change occurs, and $L_{v}$ is the vortical length scale. Both $L$ and $L_{v}$ are of the same order since any changes in the concentration from stirring occur only where vorticity is present. To derive the vortical time scale, we use the vorticity equation with only baroclinic term
\begin{equation}
    \frac{\partial \boldsymbol{\omega}}{\partial t} = \frac{\bnabla p \times \bnabla \rho}{\rho^{2}},
\end{equation}
from which we obtain,
\begin{equation}
    \frac{\ell}{c_{m}} \frac{1}{\tau_{v}} \sim \frac{\left(\frac{\Delta p}{\eta}\right) \hspace{0.8mm} \left(\frac{\Delta \rho}{\delta}\right)}{\rho_m^{2}},
    \label{eq: vortical time scale result}
\end{equation}
where $\ell/c_{m}$ is the time scale of the passage of a shock wave and $\rho_{m}$ is the total density of the mixture. The pressure jump $\Delta p$ is estimated using the mean Mach number of the shock waves and $\Delta \rho$ is a function of the Atwood number. In a similar fashion, we derive the time scale of the nonlinear dissipation term from
\begin{equation}
\frac{\partial Y_{k}}{\partial t} = \frac{D_{2}}{\mathrm{Re_{ac}}\mathrm{Sc}} \left(\bnabla Y_{k} \cdot \bnabla Y_{k} \right).
\end{equation}
The time scale of the nonlinear dissipation is obtained as, 
 \begin{equation}
    \frac{1}{\tau_{d}} \sim   \frac{|D_{2}|}{\mathrm{Re_{ac}}\mathrm{Sc} L^2}.
    \label{eq : time scale of mass fraction dissipation  }
\end{equation}
Using \eqref{eq: time scale of stirring result }, \eqref{eq: vortical time scale result}, and \eqref{eq : time scale of mass fraction dissipation  } we get a comparison of the time scales of nonlinear dissipation and  stirring as 

\begin{equation}
   \frac{\tau_{d}}{\tau_{s}} \sim \frac{ \mathrm{Re_{ac}}\mathrm{Sc} L^{2}}{|D_{2}|} \frac{\rho_m^{2}}{\left(\frac{\Delta p}{\eta}\right) \hspace{0.8mm} \left(\frac{\Delta \rho}{\delta}\right) } \frac{c_{m}}{\ell} .
\end{equation}

\begin{figure}\centering
\includegraphics[width=1.0\textwidth]{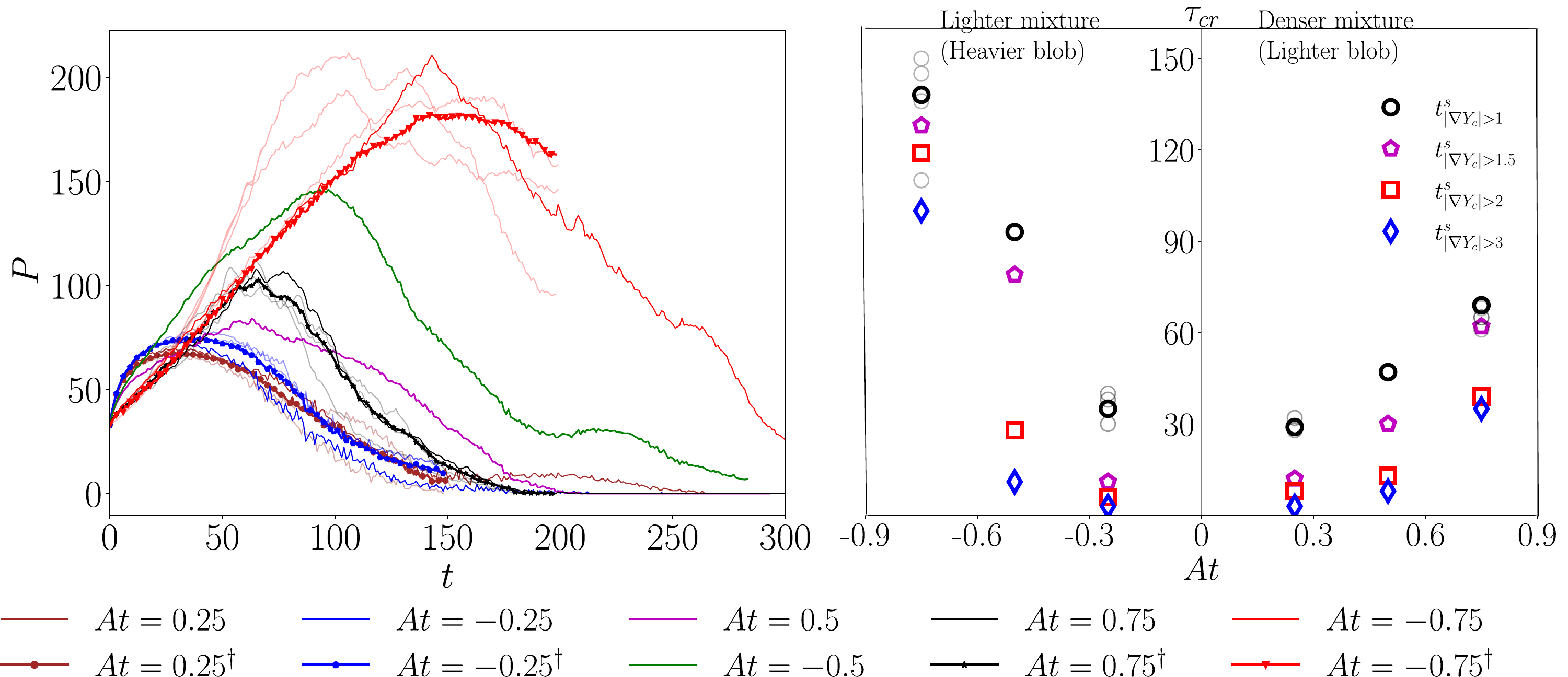}
\put(-370,165){(a)}
\put(-180,165){(b)}
\caption{\Revision{}{Evolution of the apparent interface perimeter obtained by using the condition $|\bnabla Y_{c}| > 1.0 $ (a) and the critical time of occurrence of maximum value of perimeter versus Atwood numbers (b). Shock waves stretch the interface across which molecular diffusion occurs, allowing more molecular diffusion to occur. The superscript $s$ stands for the maximum time taken to reach peak perimeter for the simulations by stirring, and the subscripts indicates the value of $\bnabla Y_{c}$ used to track the perimeter. The trend of the critical time is independent of the choice of the limiting value of $\bnabla Y_{c}$ used for calculating the interface.}}
\label{fig:Perimeter time series and tmax vs density ratio}
\end{figure}
Figure \ref{fig: Area averaged baroclinic vorticity}(b) shows the variation of the ratio of nonlinear dissipation to stirring time scales with concentration levels of the blob. We note that the time scale of stirring is faster for all cases of Atwood number for high concentration levels of the interface ($|\tau_{d} / \tau_{s}| > 1$ for high $Y_{c}$). However, it is considerably faster for $At>0$ mixtures (lighter blob, denser mixture) compared to the $At<0$ mixtures (denser blob, ligher mixture) at high concentrations, for which $\tau_s \sim \tau_d$. This also highlights that shocks are felt by the lighter inhomogeneities earlier than heavier inhomogeneities (also see $t=1$ vorticity contours for $At=0.75$ and $At=-0.75$ mixtures in figure~\ref{fig: Vorticity Contours At=-0.75}). The equivalent time scales of nonlinear dissipation and stirring for $At<0$ mixtures results in \Revision{}{an increased influence of } the movement of the interface due to the nonlinear dissipation terms  in the earlier stages of mixing of $At<0$ mixtures (interface expansion) compared to the $At>0$ mixtures (interface contraction). Furthermore, for $At>0$ mixtures (lighter blob, denser mixture), the stirring time scale becomes slower with dropping concentration levels, while for $At<0$ mixtures (heavier blob, lighter mixture), the stirring time becomes faster than the nonlinear dissipation time scales. This also indicates the dominant nature of the stirring action of the shock waves for a longer period of time for negative Atwood number cases.

We have shown that the baroclinicity is responsible for enhancing the mixing by stretching the interface of the blob. As mentioned earlier, it is difficult to track any definite interface to calculate the perimeter. But to show the effect of baroclinicity, we calculate a characteristic apparent perimeter ($P$) by tracking the area defined by $|\bnabla Y_{c}| > 1 $ and dividing it by the initial characteristic width ($\delta$ in \eqref{eq: forcing_injection}). Figure \ref{fig:Perimeter time series and tmax vs density ratio}(a) shows the evolution of the apparent perimeter of all the active scalar cases with shock waves. We see that the perimeter for the active scalars increases with an increase in baroclinicity. The peak value of the perimeter shifts to the right with an increase in the absolute value of the Atwood number. The shift is more for a heavier blob when compared to a lighter blob, as the increase in the perimeter is augmented by the outward expansion due to the nonlinear dissipation terms. Moreover, the time of occurrence of the maximum value of the perimeter ($\tau_{cr}$) depends on the choice of the limiting value of $|\bnabla Y_{c}|$ used to identify the interface for the apparent perimeter calculation. However, the trend of $\tau_{cr}$ remains identical for any choice of the limiting value provided the limiting value is not the characteristic value of the concentration gradient in the diffusive regime $(Y^{N}_c <1)$. Figure \ref{fig:Perimeter time series and tmax vs density ratio}(b) shows the occurrence of time at which maximum perimeter is achieved for different choices of $|\bnabla Y_{c}|$. The maximum time ($\tau_{cr}$) is labelled as the critical time as it distinguishes the time before which the action of stirring is dominant over molecular diffusion. We see that the $At<0$ mixtures (heavier blob, lighter mixture) have a longer stirring time than $At>0$ mixtures (lighter blob, denser mixture). The trend remains the same irrespective of the choice of interface parameter. The simulation results are in line with the time scale analysis, which shows that the stirring action is sustained longer for the $At<0$ mixtures. \Revision{reviwer 1 }{The current two-dimensional study isolates the mixing effects of the vorticity by baroclinicity. In three-dimensional cases, the mixing effects will be further enhanced by the vortex stretching mechanism.  }
\begin{figure}
\centering
	\includegraphics[width=0.8\textwidth]{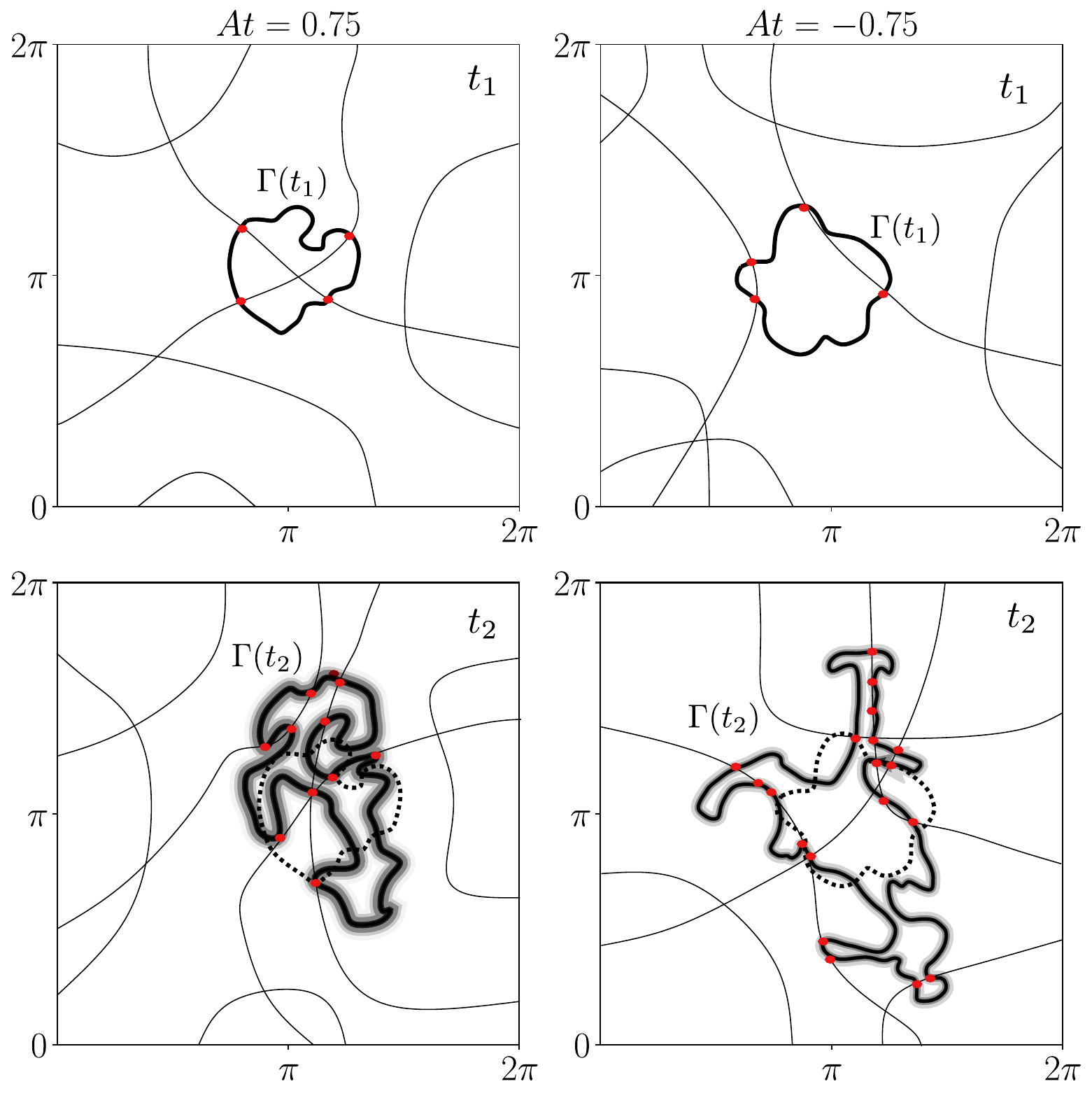}
  \put(-310,305){(a)}
\put(-155,305){(b)}
 \put(-310,153){(c)}
\put(-155,153){(d)}
	\caption{\Revision{}{Schematic illustrating the interaction of forced periodic shock waves with the concentration interface for $At=0.75$(a) and (c) and $At=-0.75$ (b) and (d) cases at time $t_1$ (a) and (b) and a later time $t_2$ (c) and (d). The shocks were traced and sketched using the dilatation values and the interface was traced and sketched using the magnitude of the concentration gradient. As the interface is stretched and folded by the baroclinic vorticity, more interaction points are introduced thus resulting in increased baroclinic interactions. This stretching is augmented by the interface expansion in the $At<0$ case due to the nonlinear dissipation term, further increasing the baroclinic interactions.}}
	\label{fig: Uncertainty schematic}
\end{figure}
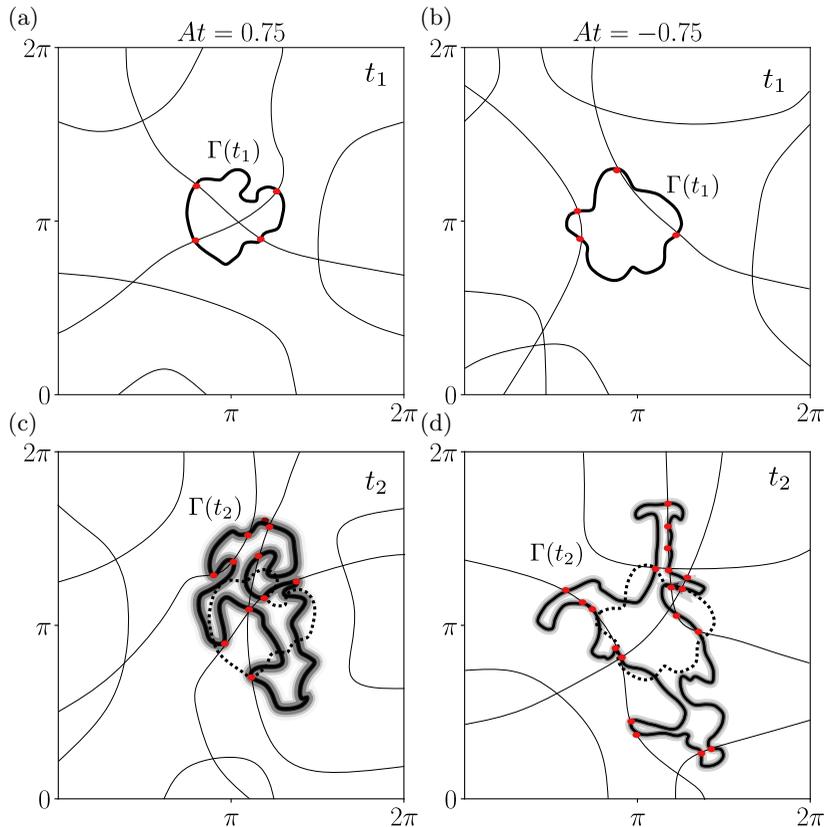

\Revision{}{In Figure \ref{fig:Perimeter time series and tmax vs density ratio}, we note that higher variation occurs in critical time values for the negative Atwood numbers stochastic realisations when compared to that of the positive Atwood numbers. The same can also be inferred from figure \ref{fig: Area averaged baroclinic vorticity}(a), where the area-averaged baroclinicity shows that the vorticity generated has a higher degree of variation for negative Atwood numbers for different stochastic realisations at the same Atwood number. We explain this difference  using the schematic shown in figure \ref{fig: Uncertainty schematic}. Figure \ref{fig: Vorticity Contours At=-0.75}(a) and (b) show that baroclinic vorticity is of significance only at the interface where the high-pressure gradients within the shock waves interact with the high-density gradients present across the interface. Figure \ref{fig: Uncertainty schematic}(a) and (b) illustrate this interaction at some initial time $t_{1}$. Figure \ref{fig: Uncertainty schematic}(a) and (b) also highlight the points of interaction between the shock waves and the density gradients across the interface. Therefore, the vorticity generated due to baroclinicity is the summation of $N$ such interactions across the interface ($\Gamma$). This can be expressed as 
\begin{equation}
    \left\langle\left|\frac{\bnabla\rho\times\bnabla p}{\rho^2}\right\rangle\right|\approx\oint\displaylimits_{\mathrm{interface}} \left|\frac{\bnabla p \times \bnabla \rho}{\rho^{2}}\right| \,d \Gamma \approx \sum_{i=1}^{N} \left| \frac{\bnabla p \times \bnabla \rho}{\rho^{2}}\right|_{i}\Delta \Gamma_i,
    \label{eq: uncertainty in baroclinicity}
\end{equation}
where the pressure gradient at each point of interaction is a stochastic process and has a variance associated with it
At a later time, $t_{2}$, the perimeter of the interface is stretched by the baroclinic vorticity for both positive and negative Atwood number cases. The same is also shown in schematic figure \ref{fig: Uncertainty schematic} (c) and (d). However, the perimeter is higher for the negative Atwood number cases compared to the positive Atwood number case due to the nonlinear dissipation term. The increase in perimeter results in more points of interaction between the pressure gradients and density gradients, resulting in more baroclinic vorticity. This indicates that for identical parameters of stochastic forcing, the negative Atwood number cases will generate more vorticity than the positive Atwood number case for the same magnitude of density gradients across the interface, and the same can also be seen in figure \ref{fig: Area averaged baroclinic vorticity} (a). This increase in  the number of interaction points also increases the uncertainty/variance associated with pressure gradients across the interface. Negative Atwood numbers have more such points when compared to positive Atwood numbers, and hence have more variation in values among realisations. But at later times, when molecular diffusion dominates mixing, the variance among realisations reduces.
}

\end{subsection}

\section{Conclusions}
\label{sec: Conclusions}

We have studied the enhancement of the mixing of an active scalar by the stirring action of stochastically forced shock waves using two-dimensional DNS. We consider the mixing of two different species of varying densities where one of the species has an initial circular profile. The Atwood number is used as the dimensionless parameter to study the effect of the shock waves on the mixing of a lighter blob ($At > 0$) and a denser blob  ($At < 0$). We simplify the diffusion term in the evolution equation for the concentration of the blob \eqref{eq: Pure Diffusion equation expanded} in terms of the concentration gradients. Comparison of the diffusion terms in concentration of passive and active scalars shows that two additional terms - concentration gradient driven \Revision{}{nonlinear} dissipation and density gradient driven \Revision{}{nonlinear} dissipation term are present in the diffusion of active scalars in addition to the Laplacian term. We show how the presence of density gradients affects mixing by considering cases where mixing is purely due to molecular diffusion. We show that the density gradients and the concentration gradients alter the coefficients of the Laplacian term and the \Revision{}{nonlinear} dissipation terms. A one-dimensional \Revision{}{unsteady nonlinear diffusion equation} is used to show that the positive coefficients for the \Revision{}{nonlinear} dissipation terms for a denser blob causes the interface to expand, while the negative coefficients which occur for a lighter blob causes the interface to shrink. This competing effect of \Revision{}{nonlinear} dissipation in active scalars \Revision{}{also} modifies the effect of shock waves \Revision{}{driven stirring} on the mixing of active scalars.

We show that the stirring action of shock waves increases the space-filling capacity of the active scalars and enhances mixing overall. Shock waves increase the mean value of concentration gradients and sustain them for longer times, allowing for more effective mixing. The lighter mixtures ($At < 0$) sustain concentration gradients longer than the denser mixtures ($At > 0$) due to the perimeter increase caused by the \Revision{}{nonlinear} dissipation terms, and the time scale of this perimeter increase being similar to the shock wave driven stirring time scale. Shock waves enhance mixing by stretching the interface across which the molecular diffusion occurs and act till all the points of the inhomogenity are exposed to the surrounding species. We give an estimate of this mixing time using the maximum concentration of the blob and show that it is also the time instant at which the area-averaged magnitude of concentration gradients peaks. For compressible flow in a two-dimensional periodic domain, baroclinicity is the only vorticity production possible, thus affecting the velocity in the advection term of the species concentration evolution equation. The interface between the active scalars exhibits large density gradients. Consequently, random shock waves interacting with this interface generate strong vorticity at the interface via the baroclinicity. Hence, a random shock field enhances the mixing of active scalars but has no effects on passive scalar mixing. We also show that the increase in density gradient increases the baroclinicity, thereby enhancing the mixing. Furthermore, shock waves do not enhance the mixing of passive scalars due to the absence of density gradients.

The current study shows that the mixing of active scalars differs significantly from that of passive scalars. \Revision{2}{In active scalar mixing, the nonlinear diffusion (termed as \Revision{}{nonlinear} dissipation in this work), the Laplacian diffusion, and the stirring interplay resulting in different mixing dynamics for heavier mixtures with lighter inhomogeneities as opposed to the lighter mixtures. The study is also a step towards a better understanding of shock-enhanced fuel-air mixing in air-breathing engines where there is very little residence time~\citep{broadwell1984structure,tian2017numerical,wong2022analysis}. The future research direction of this work is considering the effect of the vortex stretching term in sustained three-dimensional shock-resolved simulations. In three dimensions, the higher baroclinic torque in $At<0$ mixtures may result in triggering hydrodynamic turbulence, which could result in the enhacement of mixing. However, the interplay of stirring with the nonlinear diffusion of the active scalars would result in similar dynamics presented in this work. }

We acknowledge the financial support received from Science and Engineering Research Board (SERB), Government of India under Grant No. SRG/2022/000728. We also thank IIT Delhi HPC facility for computational resources.

Declaration of Interests: The authors report no conflict of interest.
\appendix
%\section{Derivation of Species Diffusion terms}\label{appA}
%\begin{subsection}{Derivation of Species Diffusion terms}
%\begin{figure}\centering
%\includegraphics[width=1.0\textwidth]{./Figures/Diff_coeff_combined}
%\put(-375,140){(a)}
%\put(-185,140){(b)}
%\caption{Variation of active scalar diffusion coefficient with concentration level of circular species
% for the laplacian term as shown in \eqref{eq: D1 coeff} . Variation of concentration-gradient driven dissipation coefficient with concentration level of circular species
% for the dissipation terms as shown in \eqref{eq: D1 coeff}.}.
%\label{fig: D1 coeff}
%\end{figure}

\section{Diffusion and dissipation time scale comparison}
\label{appB}
\begin{figure}
\centering
	\includegraphics[width=0.55\textwidth]{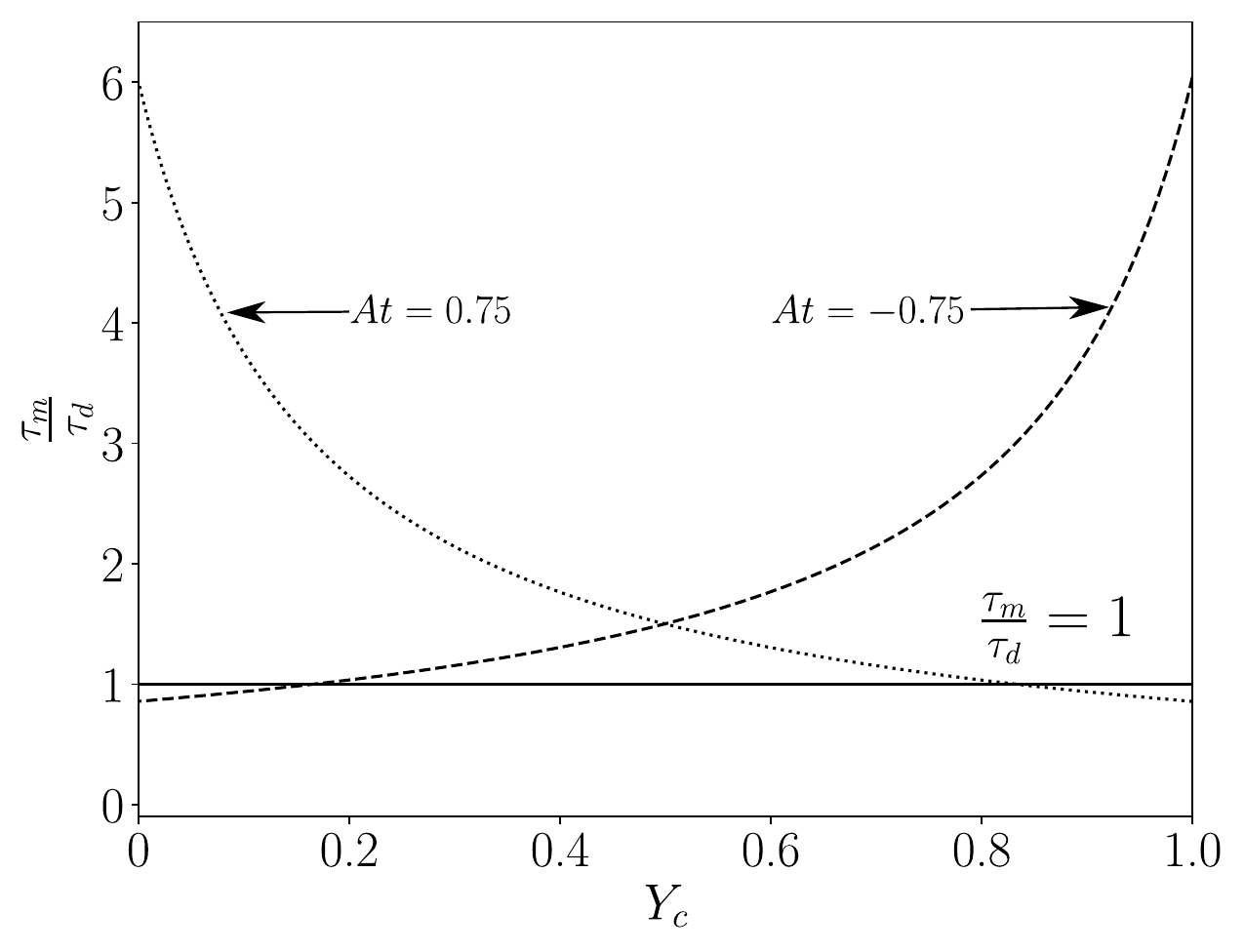}
	\caption{Variation of the ratio of Laplacian diffusion to dissipation time scales with  concentration level of blob.}
	\label{fig: time scales}
\end{figure}
In this section, we compare the time scales across which Laplacian molecular diffusion and dissipation occur.
To derive a time scale across which only molecular diffusion of active scalars driven by the Laplacian term occurs, we use the following dimensional form of the equation, 
\begin{equation}
    \frac{\partial Y_{k}}{\partial t} = \frac{D_{1}}{\mathrm{Re_{ac}}\mathrm{Sc}}\bnabla^{2}Y_{k} .
\end{equation}
Using scaling analysis, we obtain the time scale over which molecular diffusion ($\tau_{m}$) of active scalars occurs as,
\begin{equation}
   \frac{1}{\tau_{m}}  \sim \frac{D_{1}}{\mathrm{Re_{ac}}\mathrm{Sc} L^{2}}.
    \label{eq : time scale of molecular diffusion }
\end{equation}
Combining \eqref{eq : time scale of molecular diffusion } and \eqref{eq : time scale of mass fraction dissipation  }, we get the ratio of the molecular diffusion time scale ($ \tau_{m}$) to the \Revision{}{nonlinear} dissipation time scale ($ \tau_{d}$) as  
 \begin{equation}
     \frac{\tau_{m}}{\tau_{d}} \sim \frac{|D_{2}|}{D_{1}} \sim \frac{|1 - \chi|}{1 + \left( \chi -1 \right) Y_{c}}.
 \end{equation}
 Figure \ref{fig: time scales} shows the variation of the ratio of molecular diffusion time scale to \Revision{}{nonlinear} dissipation time scale with the concentration $Y_{c}$. The figure can be interpreted as the variation in time scales across the decaying moving interface. The concentration at the interface is high during the initial time, resulting in \Revision{}{nonlinear} dissipation occurring faster than molecular diffusion for $At<0$. This results in the concentration gradient decaying less and mostly resulting in a  shift/movement of the concentration gradients. But for the positive Atwood number, molecular diffusion occurs faster than \Revision{}{nonlinear} dissipation, resulting in more decay of gradients. For negative Atwood number cases, the \Revision{}{nonlinear} dissipation term dominates the molecular diffusion term, allowing higher gradients to exist longer while for positive Atwood number, molecular diffusion leads to the drop in $\langle|\bnabla Y_{c} |\rangle$. This can also be inferred from the contours plots of $|\bnabla Y_{c}|$ of figures \ref{fig:No shock nabla Y contours}(a) and \ref{fig:No shock nabla Y contours}(b).

\bibliographystyle{jfm}
\bibliography{references}

\providecommand{\noopsort}[1]{}\providecommand{\singleletter}[1]{#1}%
\begin{thebibliography}{48}
\expandafter\ifx\csname natexlab\endcsname\relax\def\natexlab#1{#1}\fi
\def\au#1{#1} \def\ed#1{#1} \def\yr#1{#1}\def\at#1{#1}\def\jt#1{\textit{#1}}
  \def\bt#1{#1}\def\bvol#1{\textbf{#1}} \def\vol#1{#1} \def\pg#1{#1}
  \def\publ#1{#1}\def\arxiv#1{#1}\def\org#1{#1}\def\st#1{\textit{#1}}

\bibitem[Abarzhi(2002)]{abarzhi2002new}
{\sc \au{Abarzhi, SI}} \yr{2002}  \at{A new type of the evolution of the bubble
  front in the richtmyer--meshkov instability}.  \jt{Physics Letters A}
  \bvol{294}~(2),  \pg{95--100}.

\bibitem[Al-Thehabey(2020)]{al2020modeling}
{\sc \au{Al-Thehabey, Omar~Yousef}} \yr{2020}  \at{Modeling the amplitude
  growth of richtmyer--meshkov instability in shock--flame interactions}.
  \jt{Physics of Fluids}  \bvol{32}~(10).

\bibitem[Arnett(2000)]{arnett2000role}
{\sc \au{Arnett, David}} \yr{2000}  \at{The role of mixing in astrophysics}.
  \jt{The Astrophysical Journal Supplement Series}  \bvol{127}~(2),  \pg{213}.

\bibitem[Atkinson {\em et~al.\/}(1981)Atkinson, Reuter \&
  Ridler-Rowe]{atkinson1981traveling}
{\sc \au{Atkinson, Colin}, \au{Reuter, Gerd~EH} \& \au{Ridler-Rowe, CJ}}
  \yr{1981}  \at{Traveling wave solution for some nonlinear diffusion
  equations}.  \jt{SIAM Journal on Mathematical Analysis}  \bvol{12}~(6),
  \pg{880--892}.

\bibitem[Augier {\em et~al.\/}(2019)Augier, Mohanan \&
  Lindborg]{Augier_Mohanan_Lindborg_2019}
{\sc \au{Augier, Pierre}, \au{Mohanan, Ashwin~Vishnu} \& \au{Lindborg, Erik}}
  \yr{2019}  \at{Shallow water wave turbulence}.  \jt{Journal of Fluid
  Mechanics}  \bvol{874},  \pg{1169–1196}.

\bibitem[Broadwell \& Breidenthal(1984)]{broadwell1984structure}
{\sc \au{Broadwell, JE} \& \au{Breidenthal, RE}} \yr{1984}  \at{Structure and
  mixing of a transverse jet in incompressible flow}.  \jt{Journal of Fluid
  Mechanics}  \bvol{148},  \pg{405--412}.

\bibitem[Brouillette(2002)]{brouillette2002richtmyer}
{\sc \au{Brouillette, Martin}} \yr{2002}  \at{The richtmyer-meshkov
  instability}.  \jt{Annual Review of Fluid Mechanics}  \bvol{34}~(1),
  \pg{445--468}.

\bibitem[Buch \& Dahm(1996)]{buch1996experimental}
{\sc \au{Buch, Kenneth~A} \& \au{Dahm, Werner~JA}} \yr{1996}  \at{Experimental
  study of the fine-scale structure of conserved scalar mixing in turbulent
  shear flows. part 1. sc [gt] 1}.  \jt{Journal of Fluid Mechanics}
  \bvol{317},  \pg{21--71}.

\bibitem[Burrows(2000)]{burrows2000supernova}
{\sc \au{Burrows, Adam}} \yr{2000}  \at{Supernova explosions in the universe}.
  \jt{Nature}  \bvol{403}~(6771),  \pg{727--733}.

\bibitem[Cho {\em et~al.\/}(2014)Cho, Venturi \&
  Karniadakis]{cho2014statistical}
{\sc \au{Cho, Heyrim}, \au{Venturi, Daniele} \& \au{Karniadakis, George~E}}
  \yr{2014}  \at{Statistical analysis and simulation of random shocks in
  stochastic burgers equation}.  \jt{Proceedings of the Royal Society A:
  Mathematical, Physical and Engineering Sciences}  \bvol{470}~(2171),
  \pg{20140080}.

\bibitem[Dimotakis(2005)]{dimotakis2005turbulent}
{\sc \au{Dimotakis, Paul~E}} \yr{2005}  \at{Turbulent mixing}.  \jt{Annu. Rev.
  Fluid Mech.}  \bvol{37},  \pg{329--356}.

\bibitem[Donzis(2012)]{donzis2012shock}
{\sc \au{Donzis, Diego~A}} \yr{2012}  \at{Shock structure in shock-turbulence
  interactions}.  \jt{Physics of Fluids}  \bvol{24}~(12).

\bibitem[Eckart(1948)]{eckart1948analysis}
{\sc \au{Eckart, Carl}} \yr{1948}  \at{An analysis of the stirring and mixing
  processes in incompressible fluids}.  \jt{Journal of Marine Research} .

\bibitem[Eswaran \& Pope(1988)]{eswaran1988examination}
{\sc \au{Eswaran, Vinayak} \& \au{Pope, Stephen~B}} \yr{1988}  \at{An
  examination of forcing in direct numerical simulations of turbulence}.
  \jt{Computers \& Fluids}  \bvol{16}~(3),  \pg{257--278}.

\bibitem[Gao {\em et~al.\/}(2016)Gao, Zhang, He \& Tian]{gao2016formula}
{\sc \au{Gao, Fujie}, \au{Zhang, Yousheng}, \au{He, Zhiwei} \& \au{Tian,
  Baolin}} \yr{2016}  \at{Formula for growth rate of mixing width applied to
  richtmyer-meshkov instability}.  \jt{Physics of Fluids}  \bvol{28}~(11).

\bibitem[Gao {\em et~al.\/}(2020)Gao, Bermejo-Moreno \&
  Larsson]{gao2020parametric}
{\sc \au{Gao, Xiangyu}, \au{Bermejo-Moreno, Ivan} \& \au{Larsson, Johan}}
  \yr{2020}  \at{Parametric numerical study of passive scalar mixing in shock
  turbulence interaction}.  \jt{Journal of Fluid Mechanics}  \bvol{895},
  \pg{A21}.

\bibitem[Gupta \& Scalo(2018)]{gupta2018spectral}
{\sc \au{Gupta, Prateek} \& \au{Scalo, Carlo}} \yr{2018}  \at{Spectral energy
  cascade and decay in nonlinear acoustic waves}.  \jt{Physical Review E}
  \bvol{98}~(3),  \pg{033117}.

\bibitem[Herrmann {\em et~al.\/}(2008)Herrmann, Moin \&
  Abarzhi]{herrmann2008nonlinear}
{\sc \au{Herrmann, Marcus}, \au{Moin, Parviz} \& \au{Abarzhi, Snezhana~I}}
  \yr{2008}  \at{Nonlinear evolution of the richtmyer--meshkov instability}.
  \jt{Journal of Fluid Mechanics}  \bvol{612},  \pg{311--338}.

\bibitem[Hirschfelder {\em et~al.\/}(1964)Hirschfelder, Curtiss \&
  Bird]{hirschfelder1964molecular}
{\sc \au{Hirschfelder, Joseph~O}, \au{Curtiss, Charles~F} \& \au{Bird,
  R~Byron}} \yr{1964} {\em The molecular theory of gases and liquids\/}.
  \publ{John Wiley \& Sons}.

\bibitem[John {\em et~al.\/}(2019)John, Donzis \&
  Sreenivasan]{john2019solenoidal}
{\sc \au{John, John~Panickacheril}, \au{Donzis, Diego~A} \& \au{Sreenivasan,
  Katepalli~R}} \yr{2019}  \at{Solenoidal scaling laws for compressible
  mixing}.  \jt{Physical Review Letters}  \bvol{123}~(22),  \pg{224501}.

\bibitem[Jossy \& Gupta(2023)]{jossy2023baroclinic}
{\sc \au{Jossy, Joaquim~P} \& \au{Gupta, Prateek}} \yr{2023}  \at{Baroclinic
  interaction of forced shock waves with random thermal gradients}.
  \jt{Physics of Fluids}  \bvol{35}~(5).

\bibitem[Kumar {\em et~al.\/}(2005)Kumar, Orlicz, Tomkins, Goodenough,
  Prestridge, Vorobieff \& Benjamin]{kumar2005stretching}
{\sc \au{Kumar, S}, \au{Orlicz, G}, \au{Tomkins, C}, \au{Goodenough, C},
  \au{Prestridge, K}, \au{Vorobieff, P} \& \au{Benjamin, R}} \yr{2005}
  \at{Stretching of material lines in shock-accelerated gaseous flows}.
  \jt{Physics of Fluids}  \bvol{17}~(8).

\bibitem[Kundu {\em et~al.\/}(2015)Kundu, Cohen \& Dowling]{kundu2015fluid}
{\sc \au{Kundu, Pijush~K}, \au{Cohen, Ira~M} \& \au{Dowling, David~R}}
  \yr{2015} {\em Fluid mechanics\/}.  \publ{Academic press}.

\bibitem[Larsson \& Lele(2009)]{larsson2009direct}
{\sc \au{Larsson, Johan} \& \au{Lele, Sanjiva~K}} \yr{2009}  \at{Direct
  numerical simulation of canonical shock/turbulence interaction}.  \jt{Physics
  of fluids}  \bvol{21}~(12).

\bibitem[Li {\em et~al.\/}(2021)Li, Tian, He \& Zhang]{li2021growth}
{\sc \au{Li, Haifeng}, \au{Tian, Baolin}, \au{He, Zhiwei} \& \au{Zhang,
  Yousheng}} \yr{2021}  \at{Growth mechanism of interfacial fluid-mixing width
  induced by successive nonlinear wave interactions}.  \jt{Physical Review E}
  \bvol{103}~(5),  \pg{053109}.

\bibitem[Liepmann \& Roshko(2001)]{liepmann2001elements}
{\sc \au{Liepmann, Hans~Wolfgang} \& \au{Roshko, Anatol}} \yr{2001} {\em
  Elements of gasdynamics\/}.  \publ{Courier Corporation}.

\bibitem[Liu {\em et~al.\/}(2022)Liu, Yu, Zhang \& Xiang]{liu2022mixing}
{\sc \au{Liu, Hong}, \au{Yu, Bin}, \au{Zhang, Bin} \& \au{Xiang, Yang}}
  \yr{2022}  \at{On mixing enhancement by secondary baroclinic vorticity in a
  shock--bubble interaction}.  \jt{Journal of Fluid Mechanics}  \bvol{931},
  \pg{A17}.

\bibitem[Meshkov(1969)]{meshkov1969instability}
{\sc \au{Meshkov, \_E~E}} \yr{1969}  \at{Instability of the interface of two
  gases accelerated by a shock wave}.  \jt{Fluid Dynamics}  \bvol{4}~(5),
  \pg{101--104}.

\bibitem[Meunier \& Villermaux(2003)]{meunier2003vortices}
{\sc \au{Meunier, Patrice} \& \au{Villermaux, Emmanuel}} \yr{2003}  \at{How
  vortices mix}.  \jt{Journal of Fluid Mechanics}  \bvol{476},  \pg{213--222}.

\bibitem[Miura \& Kida(1995)]{miura1995acoustic}
{\sc \au{Miura, Hideaki} \& \au{Kida, Shigeo}} \yr{1995}  \at{Acoustic energy
  exchange in compressible turbulence}.  \jt{Physics of Fluids}  \bvol{7}~(7),
  \pg{1732--1742}.

\bibitem[Ni(2016)]{ni2016compressible}
{\sc \au{Ni, Qionglin}} \yr{2016}  \at{Compressible turbulent mixing: Effects
  of compressibility}.  \jt{Physical Review E}  \bvol{93}~(4),  \pg{043116}.

\bibitem[Poinsot \& Veynante(2005)]{poinsot2005theoretical}
{\sc \au{Poinsot, Thierry} \& \au{Veynante, Denis}} \yr{2005} {\em Theoretical
  and numerical combustion\/}.  \publ{RT Edwards, Inc.}

\bibitem[Richtmyer(1954)]{richtmyer1954taylor}
{\sc \au{Richtmyer, Robert~D}} \yr{1954}  \bt{Taylor instability in shock
  acceleration of compressible fluids}. {\em Tech. Rep.\/}.  \org{Los Alamos
  Scientific Lab., N. Mex.}

\bibitem[Samtaney {\em et~al.\/}(2001)Samtaney, Pullin \&
  Kosovi{\'c}]{samtaney2001direct}
{\sc \au{Samtaney, Ravi}, \au{Pullin, Dale~I} \& \au{Kosovi{\'c}, Branko}}
  \yr{2001}  \at{Direct numerical simulation of decaying compressible
  turbulence and shocklet statistics}.  \jt{Physics of Fluids}  \bvol{13}~(5),
  \pg{1415--1430}.

\bibitem[Schumacher \& Sreenivasan(2005)]{schumacher2005statistics}
{\sc \au{Schumacher, J{\"o}rg} \& \au{Sreenivasan, Katepalli~R}} \yr{2005}
  \at{Statistics and geometry of passive scalars in turbulence}.  \jt{Physics
  of Fluids}  \bvol{17}~(12).

\bibitem[Schumacher {\em et~al.\/}(2005)Schumacher, Sreenivasan \&
  Yeung]{schumacher2005very}
{\sc \au{Schumacher, Joerg}, \au{Sreenivasan, Katepalli~R} \& \au{Yeung, PK}}
  \yr{2005}  \at{Very fine structures in scalar mixing}.  \jt{Journal of Fluid
  Mechanics}  \bvol{531},  \pg{113--122}.

\bibitem[Sreenivasan(2019)]{sreenivasan2019turbulent}
{\sc \au{Sreenivasan, Katepalli~R}} \yr{2019}  \at{Turbulent mixing: A
  perspective}.  \jt{Proceedings of the National Academy of Sciences}
  \bvol{116}~(37),  \pg{18175--18183}.

\bibitem[Tennekes \& Lumley(1972)]{tennekes1972first}
{\sc \au{Tennekes, Hendrik} \& \au{Lumley, John~Leask}} \yr{1972} {\em A first
  course in turbulence\/}.  \publ{MIT press}.

\bibitem[Thomas \& Kares(2012)]{thomas2012drive}
{\sc \au{Thomas, Vincent~A} \& \au{Kares, Robert~J}} \yr{2012}  \at{Drive
  asymmetry and the origin of turbulence in an icf implosion}.  \jt{Physical
  review letters}  \bvol{109}~(7),  \pg{075004}.

\bibitem[Tian {\em et~al.\/}(2017)Tian, Jaberi, Li \&
  Livescu]{tian2017numerical}
{\sc \au{Tian, Yifeng}, \au{Jaberi, Farhad~A}, \au{Li, Zhaorui} \& \au{Livescu,
  Daniel}} \yr{2017}  \at{Numerical study of variable density turbulence
  interaction with a normal shock wave}.  \jt{Journal of Fluid Mechanics}
  \bvol{829},  \pg{551--588}.

\bibitem[Tomkins {\em et~al.\/}(2008)Tomkins, Kumar, Orlicz \&
  Prestridge]{tomkins2008experimental}
{\sc \au{Tomkins, C}, \au{Kumar, S}, \au{Orlicz, G} \& \au{Prestridge, K}}
  \yr{2008}  \at{An experimental investigation of mixing mechanisms in
  shock-accelerated flow}.  \jt{Journal of Fluid Mechanics}  \bvol{611},
  \pg{131--150}.

\bibitem[Villermaux(2019)]{villermaux2019mixing}
{\sc \au{Villermaux, Emmanuel}} \yr{2019}  \at{Mixing versus stirring}.
  \jt{Annual Review of Fluid Mechanics}  \bvol{51},  \pg{245--273}.

\bibitem[Wong {\em et~al.\/}(2022)Wong, Baltzer, Livescu \&
  Lele]{wong2022analysis}
{\sc \au{Wong, Man~Long}, \au{Baltzer, Jon~R}, \au{Livescu, Daniel} \&
  \au{Lele, Sanjiva~K}} \yr{2022}  \at{Analysis of second moments and their
  budgets for richtmyer-meshkov instability and variable-density turbulence
  induced by reshock}.  \jt{Physical Review Fluids}  \bvol{7}~(4),
  \pg{044602}.

\bibitem[Yang {\em et~al.\/}(1993)Yang, Kubota \&
  Zukoski]{yang1993applications}
{\sc \au{Yang, Joseph}, \au{Kubota, Toshi} \& \au{Zukoski, Edward~E}} \yr{1993}
   \at{Applications of shock-induced mixing to supersonic combustion}.
  \jt{AIAA journal}  \bvol{31}~(5),  \pg{854--862}.

\bibitem[Yu {\em et~al.\/}(2021)Yu, Liu \& Liu]{yu2021scaling}
{\sc \au{Yu, Bin}, \au{Liu, Haoyang} \& \au{Liu, Hong}} \yr{2021}  \at{Scaling
  behavior of density gradient accelerated mixing rate in shock bubble
  interaction}.  \jt{Physical Review Fluids}  \bvol{6}~(6),  \pg{064502}.

\bibitem[Yuan {\em et~al.\/}(2023)Yuan, Zhao, Liu, Wang, Liu \&
  Lu]{yuan2023instability}
{\sc \au{Yuan, Ming}, \au{Zhao, Zhiye}, \au{Liu, Luoqin}, \au{Wang, Pei},
  \au{Liu, Nan-Sheng} \& \au{Lu, Xi-Yun}} \yr{2023}  \at{Instability evolution
  of a shock-accelerated thin heavy fluid layer in cylindrical geometry}.
  \jt{Journal of Fluid Mechanics}  \bvol{969},  \pg{A6}.

\bibitem[Zhang(1998)]{zhang1998analytical}
{\sc \au{Zhang, Qiang}} \yr{1998}  \at{Analytical solutions of layzer-type
  approach to unstable interfacial fluid mixing}.  \jt{Physical review letters}
   \bvol{81}~(16),  \pg{3391}.

\bibitem[Zhou(2017)]{zhou2017rayleigh}
{\sc \au{Zhou, Ye}} \yr{2017}  \at{Rayleigh--taylor and richtmyer--meshkov
  instability induced flow, turbulence, and mixing. ii}.  \jt{Physics Reports}
  \bvol{723},  \pg{1--160}.

\end{thebibliography}

\end{document}